\def\ARXIV{} % enable real authors in main
\def\ARXIV{} % enable real authors in main
\documentclass[letterpaper,twocolumn,10pt]{article}
\usepackage{usenix2019_v3}

\newcommand{\eg}{{\it e.g.,\ }}

\newcommand{\aka}{{\it a.k.a.\ }}

\usepackage{comment}
\usepackage[english]{babel}
\usepackage{graphicx}
\usepackage{xspace}
\usepackage{pifont}
\usepackage{xcolor}
\usepackage{blindtext}
\usepackage{footnote}
\usepackage[noend]{algpseudocode}
\usepackage{algorithm}

\usepackage{amsmath}
\usepackage{cleveref}
\crefname{section}{}{\S\S}
\crefdefaultlabelformat{\S#2#1#3}
\usepackage{caption}
\usepackage{multirow} 
\usepackage{subcaption}
\usepackage{amsfonts}
\usepackage{stmaryrd}
\usepackage{msc}

\newcommand{\jjh}[1]{\textcolor{blue}{[jjh: #1]}}

\newcommand{\para}[1]{\noindent {\bf #1}}

\ifdefined\ARXIV
\newcommand{\system}{Argo\xspace}
\else
\newcommand{\system}{Procurator\xspace}
\fi

\date{}

\begin{document}
\ifdefined\ARXIV
\title{Argo: An efficient verification framework for distributed in-network computing programs}
\else
\title{Toward Efficient Verification of Distributed In-Network Computing Programs}
\fi

% Author/affiliation block controlled by ARXIV flag
\ifdefined\ARXIV
  % Real author block for arXiv build
% Fill in real names/affiliations/emails as appropriate
\author{%
  \rm Mingyuan Song$^{1}$,
  \rm Huan Shen$^{1}$,
  \rm Jinghui Jiang$^{1}$,
  \rm Qiang Su$^{1}$,
  \rm Qingyu Song$^{1}$,
  \rm Lu Tang$^{1}$\\
  \rm Wanjian Feng$^{2}$,
  \rm Fei Yuan$^{1}$,
  \rm Qiao Xiang$^{1}$,
  \rm Jiwu Shu$^{1}$\\[0.4em]
  $^{1}$Xiamen University \quad $^{2}$Yealink
}

% \texttt{mingyuan.song.xmu@gmail.com, shenhuan@stu.xmu.edu.cn, jinghuijiang314@gmail.com, qiangsu@xmu.edu.cn, xiangq27@gmail.com}
% \date{}

\else
  % Anonymous author block for blind submission build
\author{Paper \# 1047, 12 pages body, 3 pages reference, 2 pages appendix}

\fi

\maketitle

\ifdefined\ARXIV

\begin{abstract}

Distributed in-network programs are increasingly deployed in data centers for their performance benefits, but shifting application logic to switches also enlarges the failure domain. Ensuring their correctness before deployment is thus critical for reliability.
While prior verification frameworks can efficiently detect bugs for programs running on a single switch, they overlook the common interactive behaviors in distributed settings, 
thereby missing related bugs that can cause state inconsistencies and system failures.

This paper presents \system, a verification framework that efficiently captures interactive behaviors in distributed in-network programs. \system introduces a formal model combining the actor paradigm with Communicating Sequential Processes (CSP), translating pipeline execution into reactive, event-driven actors and unifying their interactions as message passing. To support flexible specification of distributed properties, it provides a unified intent language. Additionally, it incorporates a semantic-aware state pruner to reduce verification complexity, thus ensuring system scalability.
Evaluation results show that \system efficiently uncovers 10 distinct bugs caused by interactive behaviors across five real-world in-network systems. It also reduces verification time by up to 913.2$\times$ and memory consumption by up to 1.9$\times$ compared to the state-of-the-art verifier.

\end{abstract}
\else

\begin{abstract}

Distributed in-network programs are increasingly deployed in data centers for their performance benefits, but shifting application logic to switches also enlarges the failure domain. Ensuring their correctness before deployment is thus critical for reliability.
While prior verification frameworks can efficiently detect bugs for programs running on a single switch, they overlook the common interactive behaviors in distributed settings, 
thereby missing related bugs that can cause state inconsistencies and system failures.

This paper presents \system, a verification framework that efficiently captures interactive behaviors in distributed in-network programs. \system introduces a formal model combining the actor paradigm with Communicating Sequential Processes (CSP), translating pipeline execution into reactive, event-driven actors and unifying their interactions as message passing. To support flexible specification of distributed properties, it provides a unified intent language. Additionally, it incorporates a semantic-aware state pruner to reduce verification complexity, thus ensuring system scalability.
Evaluation results show that \system efficiently uncovers 10 distinct bugs caused by interactive behaviors across five real-world in-network systems. It also reduces verification time by up to 913.2$\times$ and memory consumption by up to 1.9$\times$ compared to the state-of-the-art verifier.

\end{abstract}
\fi
\ifdefined\ARXIV
\section{Introduction}\label{sec:intro}

Modern data centers increasingly deploy in-network programs on programmable switches to accelerate distributed systems. These programs are typically written in P4\cite{p4} for its flexibility and efficiency in packet processing. They often maintain application state locally while interacting with backend servers or coordinating with peer switches when deployed in a distributed manner~\cite{netchain,xos,li2022switchtx,netlock,p4db,concordia,sheng2023farreach,atp,netcache,distcache,SwiSh}. 
As in-network programs grow in feature coverage and cross-device coordination, their complexity and hence the potential risk surface also increase~\cite {li2022cetus,neves2019p4box,ruffy2020gauntlet}, such as misforwarding, unintended loops, or compiler/firmware-induced misbehaviors~\cite{nvd-cve-2024-0106,ruffy2020gauntlet,neves2019p4box}.

%However, In stateful in-network systems, dataplane programmability introduces new failure and attack surfaces: a single logical mistake can cascade at line rate across devices and manifest as production incidents (e.g., misforwarding, unintended loops, or compiler/firmware-induced misbehavior)~\cite{nvd-cve-2024-0106,ruffy2020gauntlet,neves2019p4box}. 

In practice, ensuring correct behaviors of distributed in-network programs is critical, but relying on human expertise to check the programs is both unscalable and error-prone. To address this, numerous automated verification tools have been developed to detect potential bugs and faults before deployment~\cite{p4v,bf4,vera,p4inv,p4tv,assert-p4-conext,assert-p4-sosr}.
Existing work often treats distributed in-network programs as individual P4 instances.
Some model the program execution as an instruction pipeline, but cannot capture the state update behaviors in typical distributed in-network programs~\cite{p4v,bf4,vera,assert-p4-conext,assert-p4-sosr}. 
More recent approaches extend this to a while-loop over the pipeline, enabling state tracking during verification~\cite{p4inv,p4tv}.
While effective for detecting single-program bugs, this abstraction fails to capture switch–server and switch–switch interactions in real deployments --- the behaviors between any two programs cannot be described by a single while-loop-based modeling. As a result, the subtle bugs from the interactive behaviors are left unnoticed, leading to failure risks like state inconsistency and misforwarding.
%\qs{what problems does it cause?}

The fundamental question is, how can we build an efficient verification framework that can detect subtle bugs arising from interactive behaviors in real-world distributed in-network programs?

To address this question, we make two key observations on distributed in-network program execution. 
First, a switch program behaves as a long-running process, triggered by incoming packets, performing local stateful actions, and generating output packets to interact with other switches or servers. These behaviors logically align with the actor paradigm in distributed systems~\cite{actor-origin}.
Second, interactions between switch programs correspond to inter-process communication, whose properties, such as ordering and loss, can be naturally captured by the classical Communication Sequential Process (CSP) model~\cite{hoare1978communicating}. 
Therefore, unlike the while-loop-based models of prior work, we propose to model distributed in-network programs as actors and their interactions as CSP.

However, realizing this idea presents several challenges.
First, distributed in-network programs feature complex P4 semantics (\eg recirculation, mirroring, state registers) that standard actor or CSP models cannot capture. To address this, we introduce a formal semantic translation that models each pipeline as a reactive event-driven actor and unifies actor interactions through message passing.
This formalization produces a model suitable for verification using classical model checkers (\eg SPIN)~\cite{spin}.
Second, complex actor interactions produce a vast number of state updates, making it difficult to bound verification overhead to ensure scalability. We tackle this with a Semantic-Aware State Pruner that applies a tailored backward slicing algorithm. This algorithm correctly models architectural loops while aggressively pruning irrelevant states and execution paths, thereby ensuring verification efficiency.
Third, existing verification intent specifications lack the flexibility to express both distributed properties (\eg consistency) and interaction topologies. We overcome this by providing a unified intent language that lets users define the network topology, state local assertions within devices, and express global invariants across devices in Linear Temporal Logic.

Specifically, we build a new verification framework called \system following this design.
It takes as input a set of distributed in-network programs along with a verification specification written in \system’s Unified Intent Language. First, it constructs a program dependency graph and applies the Semantic-Aware State Pruner to optimize state spaces, thus generating an optimized intermediate representation. 
This representation is then processed by the Semantic Translation Model, which converts it into a unified actor-and-CSP-based model that can be verified by a classical model checker (\eg SPIN). Finally, the distributed in-network program verification is completed.

We implemented a prototype of \system and will release it upon the publication of this manuscript. It consists of $\sim$12k lines of C++ and $\sim$2k lines of Python, and supports the V1Model and TNA in-network programming frameworks for software and Tofino switches. The prototype integrates with the standard P4 toolchain via the p4c front-end compiler and leverages the SPIN model checker for backend verification. 
We conduct extensive experiments on five real-world distributed in-network systems with interactive behaviors, as well as four single-device in-network programs. The results show that \system successfully uncovers 10 distinct bugs arising from interactive behaviors. For single-device programs, it matches the bug-detection capability of the state-of-the-art tool p4tv\cite{p4tv} while reducing verification time by up to 913.2$\times$ and memory consumption by up to 1.9$\times$.

\else
\section{Introduction}\label{sec:intro}

Modern data centers increasingly deploy in-network programs on programmable switches to accelerate distributed systems. These programs are typically written in P4\cite{p4} for its flexibility and efficiency in packet processing. They often maintain application state locally while interacting with backend servers or coordinating with peer switches when deployed in a distributed manner~\cite{netchain,xos,li2022switchtx,netlock,p4db,concordia,sheng2023farreach,atp,netcache,distcache,SwiSh}. 
As in-network programs grow in feature coverage and cross-device coordination, their complexity and hence the potential risk surface also increase~\cite {li2022cetus,neves2019p4box,ruffy2020gauntlet}, such as misforwarding, unintended loops, or compiler/firmware-induced misbehaviors~\cite{nvd-cve-2024-0106,ruffy2020gauntlet,neves2019p4box}.

%However, In stateful in-network systems, dataplane programmability introduces new failure and attack surfaces: a single logical mistake can cascade at line rate across devices and manifest as production incidents (e.g., misforwarding, unintended loops, or compiler/firmware-induced misbehavior)~\cite{nvd-cve-2024-0106,ruffy2020gauntlet,neves2019p4box}. 

In practice, ensuring correct behaviors of distributed in-network programs is critical, but relying on human expertise to check the programs is both unscalable and error-prone. To address this, numerous automated verification tools have been developed to detect potential bugs and faults before deployment~\cite{p4v,bf4,vera,p4inv,p4tv,assert-p4-conext,assert-p4-sosr}.
Existing work often treats distributed in-network programs as individual P4 instances.
Some model the program execution as an instruction pipeline, but cannot capture the state update behaviors in typical distributed in-network programs~\cite{p4v,bf4,vera,assert-p4-conext,assert-p4-sosr}. 
More recent approaches extend this to a while-loop over the pipeline, enabling state tracking during verification~\cite{p4inv,p4tv}.
While effective for detecting single-program bugs, this abstraction fails to capture switch–server and switch–switch interactions in real deployments --- the behaviors between any two programs cannot be described by a single while-loop-based modeling. As a result, the subtle bugs from the interactive behaviors are left unnoticed, leading to failure risks like state inconsistency and misforwarding.
%\qs{what problems does it cause?}

The fundamental question is, how can we build an efficient verification framework that can detect subtle bugs arising from interactive behaviors in real-world distributed in-network programs?

To address this question, we make two key observations on distributed in-network program execution. 
First, a switch program behaves as a long-running process, triggered by incoming packets, performing local stateful actions, and generating output packets to interact with other switches or servers. These behaviors logically align with the actor paradigm in distributed systems~\cite{actor-origin}.
Second, interactions between switch programs correspond to inter-process communication, whose properties, such as ordering and loss, can be naturally captured by the classical Communication Sequential Process (CSP) model~\cite{hoare1978communicating}. 
Therefore, unlike the while-loop-based models of prior work, we propose to model distributed in-network programs as actors and their interactions as CSP.

However, realizing this idea presents several challenges.
First, distributed in-network programs feature complex P4 semantics (\eg recirculation, mirroring, state registers) that standard actor or CSP models cannot capture. To address this, we introduce a formal semantic translation that models each pipeline as a reactive event-driven actor and unifies actor interactions through message passing.
This formalization produces a model suitable for verification using classical model checkers (\eg SPIN)~\cite{spin}.
Second, complex actor interactions produce a vast number of state updates, making it difficult to bound verification overhead to ensure scalability. We tackle this with a Semantic-Aware State Pruner that applies a tailored backward slicing algorithm. This algorithm correctly models architectural loops while aggressively pruning irrelevant states and execution paths, thereby ensuring verification efficiency.
Third, existing verification intent specifications lack the flexibility to express both distributed properties (\eg consistency) and interaction topologies. We overcome this by providing a unified intent language that lets users define the network topology, state local assertions within devices, and express global invariants across devices in Linear Temporal Logic.

Specifically, we build a new verification framework called \system following this design.
It takes as input a set of distributed in-network programs along with a verification specification written in \system’s Unified Intent Language. First, it constructs a program dependency graph and applies the Semantic-Aware State Pruner to optimize state spaces, thus generating an optimized intermediate representation. 
This representation is then processed by the Semantic Translation Model, which converts it into a unified actor-and-CSP-based model that can be verified by a classical model checker (\eg SPIN). Finally, the distributed in-network program verification is completed.

We implemented a prototype of \system and will release it upon the publication of this manuscript. It consists of $\sim$12k lines of C++ and $\sim$2k lines of Python, and supports the V1Model and TNA in-network programming frameworks for software and Tofino switches. The prototype integrates with the standard P4 toolchain via the p4c front-end compiler and leverages the SPIN model checker for backend verification. 
We conduct extensive experiments on five real-world distributed in-network systems with interactive behaviors, as well as four single-device in-network programs. The results show that \system successfully uncovers 10 distinct bugs arising from interactive behaviors. For single-device programs, it matches the bug-detection capability of the state-of-the-art tool p4tv\cite{p4tv} while reducing verification time by up to 913.2$\times$ and memory consumption by up to 1.9$\times$.

\fi
\section{Background and Motivation}

\label{sec:overview}

We start by presenting the background of verifying distributed in-network programs and analyzing the typical behaviors of these systems. We then discuss our design choice.

\subsection{Background}\label{sec:background}

\noindent{\bf Distributed in-network programs.}
With the nature of in-network and line-rate processing, programmable switches are increasingly used to offload parts of the logic of distributed systems~\cite{p4,Tofino,netcache,netchain,xos,netlock,li2022switchtx,p4db,li2020pegasus,concordia,distcache,incbricks,sheng2023farreach,atp,p4lru,gecko,clickinc,horus,lyra,switchv2p}.
For instance, Gecko~\cite{gecko} implements a P4-based load balancer on ToR switches to redirect read and write requests to distributed key-value storage servers.
In addition, in-network P4 programs can be deployed across multiple switches; for example, NetChain offloads key-value store caches to the on-chip memory of a chain of switches~\cite{netchain}. 
Such systems always rely on streaming-based interactions, either between switches and backend servers or directly among switches across the network.

\noindent{\bf P4 program verification.}
In-network P4 programs must be verified before deployment to prevent system failures caused by unexpected program errors. For stateless programs, existing verification methods typically model program execution as a pipeline of instructions, with inputs and outputs represented as mathematical symbols~\cite{p4v,bf4,aquila,p4nod,assert-p4-sosr,assert-p4-conext,vera}. The output is then verified against predefined assertions to ensure correctness. For stateful programs, recent works like P4Inv~\cite{p4inv} and p4tv~\cite{p4tv} introduce a while-loop in the instruction pipeline, iteratively updating states in hardware registers, but conduct verification based on mathematical symbols and state-machine-based model checking~\cite{modelchecking}. 

While these approaches are effective in verifying individual P4 programs, they overlook the potential errors arising from the interactive behaviors in distributed in-network programs. In the following, we showcase such issues.

\subsection{Motivating Examples}
\label{lbl:moti_exam}

In this section, we use two typical examples to illustrate the inherent error-prone behaviors in distributed in-network programs, caused by interaction between switches and backend servers or directly among switches.

\subsubsection{Switch-Server Interaction}

In many in-network systems~\cite{gecko, switcharoo, SwiSh, distcache, sheng2023farreach}, switches must redirect client requests to multiple backend servers (\aka multicast fan-out). This is typically implemented by writing P4 programs that send the first copy of a packet on its initial pass and use \textit{recirculation} to emit additional copies on subsequent passes. 
However, this mechanism can introduce inconsistent request ordering across backend servers. To illustrate this, we consider a switch $T$ and two backend key-value replicas, $S_1$ and $S_2$: $T$ delivers each write request to one server directly (\aka the first pass \texttt{p1}), where the server is chosen randomly based on load balancing (LB), and then uses recirculation to send a copy to the other (\aka the second pass \texttt{p2}).

\begin{figure}[t]
    \centering
    \includegraphics[width=0.85\linewidth]{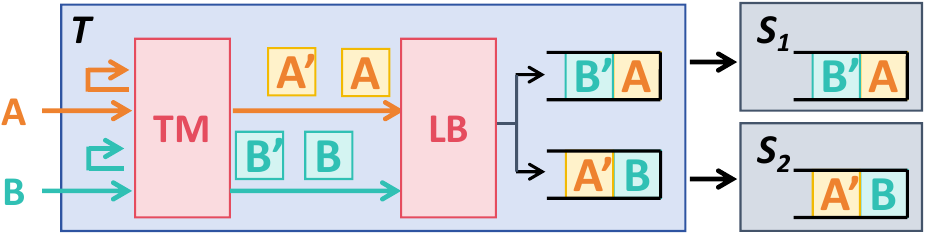}
    \caption{An example of \textit{Replica Total Order} violation.}
    \label{fig:rto}
\end{figure}

As shown in Figure~\ref{fig:rto}, two write requests, $A$ and $B$, arrive at $T$ back-to-back. When packet $A$ arrives at the traffic manager (TM), the traffic manager flags it for recirculation. However, the recirculation operation is not atomic. Therefore, TM then starts to handle packet $B$, and flags it for recirculation. Thus, LB receives data packets in the order of $A$, $B$, $A'$, and $B'$. For $A$ and $B$, the load balancer sends them to $S_1$ and $S_2$, respectively, while $A'$ and $B'$ are sent to $S_2$ and $S_1$, respectively, according to functional requirements.

\iffalse{
As shown in Figure~\ref{fig:rto}, two write requests, $A$ and $B$, arrive at $T$ back-to-back. $T$ sends $A$ to $S_1$ during \texttt{p1} and flags it for recirculation. Before $A$'s \texttt{p2}, $B$ arrives. $B$ is first sent to $S_2$ based on LB decisions and is flagged for recirculation. When $A$ enters \texttt{p2}, the recirculated $A$ re-enters the pipeline and is forwarded to $S_2$. However, during $B$'s \texttt{p2}, the recirculated $B$ is forwarded to $S_1$.
}\fi

Clearly, $S_1$ observes the order $A \rightarrow B$, while $S_2$ observes $B \rightarrow A$, which violates consistency between replicas. The root cause of this issue is architectural: recirculation splits a single logical broadcast into multiple sequential sends, which can easily lead to error-prone interleaving when multiple broadcasts coexist. While developers' expertise could help prevent such issues, this behavior still requires verification before deployment to production environments to ensure reliability.

\subsubsection{Switch-Switch Interaction}
\label{subsub:switch_inter}

\begin{figure}[t]
\centering\includegraphics[width=0.99\linewidth]{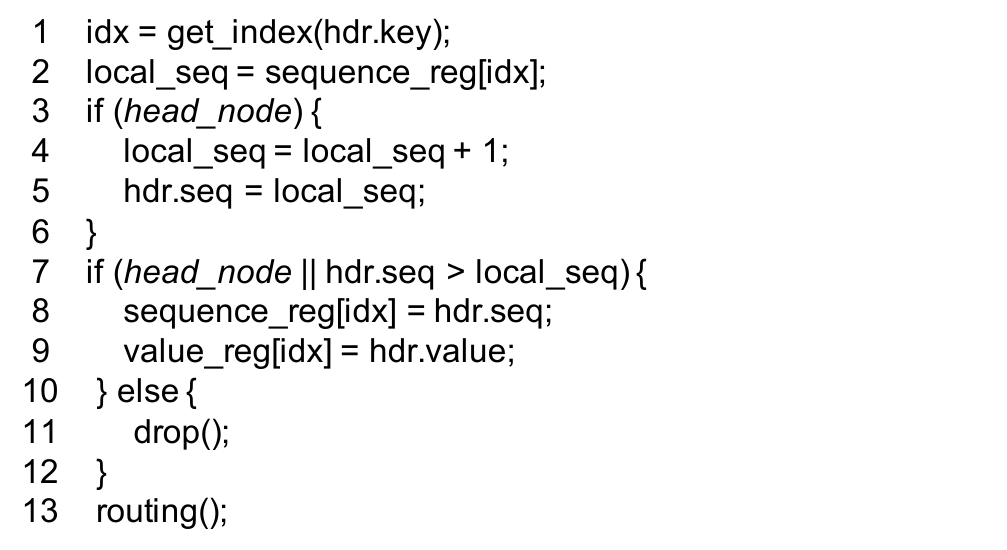}
    \caption{Code snippet for the key functionality of NetChain~\cite{netchain}. }
    \label{fig:netchain_code}
    \vspace{-5pt}
\end{figure}

\begin{figure}[t]
\centering\includegraphics[width=0.99\linewidth]{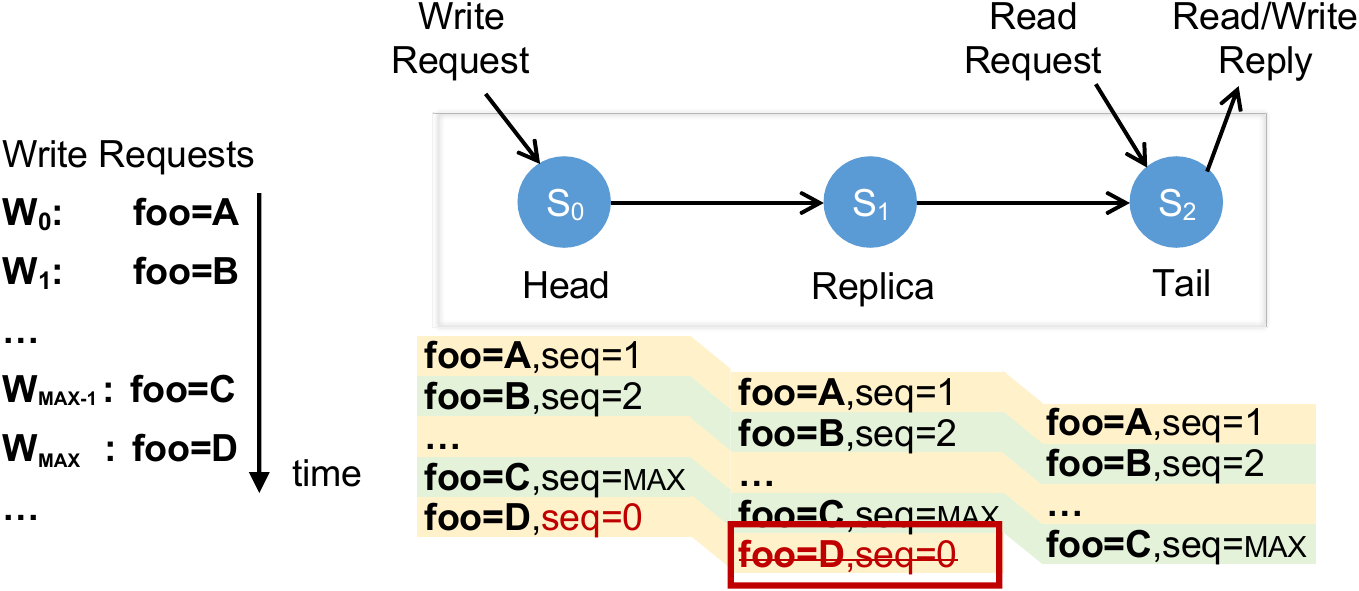}
    \caption{An example of a distributed P4 system bug. }
    \label{fig:netchain_workflow}
\end{figure}

We use NetChain~\cite{netchain} as a case study to illustrate potential errors arising from switch-switch interactions. NetChain stores cache replicas across a chain of switches in a key-value system, maintaining consistency via sequence numbers. As shown in Figure~\ref{fig:netchain_code}, when the head switch receives a write request, it increments its local sequence number, stores the data, and forwards both the data and sequence number to the subsequent switches. Each subsequent switch updates its local replica only if the incoming sequence number is larger than its local one, ensuring consistency across the chain.

NetChain’s design has been fully verified using TLA+~\cite{tlap}, yet errors can still arise from switch-to-switch interactions, often overlooked by prior verification methods. Figure~\ref{fig:netchain_workflow} shows an example of such an error.

Switches $S_1$, $S_2$, and $S_3$ form a chain, with $S_0$ as the head and $S_2$ as the tail. The system begins with no data stored, and $W_i$ denotes a write request, while $seq$ (ranging from 0 to $MAX$) tracks sequence numbers.
When $W_0$ arrives, all switches set \texttt{foo} to \texttt{A} and $seq$ to 1. Upon receiving $W_1$, $S_0$ updates \texttt{foo} to \texttt{B}, increments $seq$ to 2, and forwards the request along with the updated $seq$. Both $S_1$ and $S_2$ update \texttt{foo} to \texttt{B} and set $seq$ to 2, preserving consistency.

However, when $seq$ reaches $MAX$, the behavior starts changing. Upon receiving $W_{MAX-1}$, all switches update \texttt{foo} to \texttt{C} and set $seq$ to $MAX$. When $W_{MAX}$ arrives, $S_0$’s $seq$ overflows, wrapping around to zero (lines~3-4 in Figure~\ref{fig:netchain_code}). It forwards $W_{MAX} = D$ with $seq = 0$, but $S_1$ and $S_2$, having $seq = MAX$, ignore the update.
Therefore, the data in $S_1$ conflicts with $S_0$, and this consistency bug cannot be detected by verifying each program individually due to cross-switch interactions and version wrap-around.

\vspace{-1em}
\subsection{Our Design Choice}

In \Cref{lbl:moti_exam}, we demonstrate errors caused by switch-server, switch-switch interactions, and P4-specific features (\eg recirculation) in distributed in-network programs. This makes it urgent to design a verification approach that efficiently captures these behaviors.

A straightforward solution might be to extend existing stateful P4 verification tools (\eg P4Inv~\cite{p4inv}, p4tv~\cite{p4tv}) to account for interactive behaviors in distributed in-network programs. However, this approach is fundamentally infeasible. Existing tools treat P4 features like recirculation and mirroring as iterations within a while loop, making them incapable of detecting logical errors that occur between loops --- the root cause of the issues in \Cref{lbl:moti_exam}. Moreover, the while-loop modeling lacks visibility into external interactions (with other switches or servers), making direct extension of these tools impractical.

Therefore, we must design a new and systematic verification approach. 
Our core insight is that a distributed in-network program consists of independent, reactive computational units communicating over network channels. Based on this, we leverage two classical techniques that align well with these behaviors:

\noindent{\bf The Actor paradigm.}
A switch operates as a continuously running process, reacting to packet arrivals, performing stateful actions, and generating events (\eg sending packets). This aligns with the Actor paradigm\cite{actor-origin}, where an actor encapsulates its state and interacts via message passing. 
An actor performs operations based on the incoming message: (1) triggers a process handler to manipulate its own states; (2) interacts with others by sending messages. This approach has proven efficient in complex distributed applications\cite{newell2016optimizing, kraft2022data}, making it ideal for modeling distributed in-network programs, in contrast to the traditional sequential while-loop approach.

\noindent{\bf The Communicating Sequential Process (CSP) model.}
Each switch program is an independent process, and the packet exchanges between them constitute inter-process communication. The CSP\cite{hoare1978communicating} model is a classic framework that describes and analyzes such concurrent process interactions, allowing us to explicitly define semantics like ordering and loss, which are crucial for distributed in-network programs.

Thus, our approach, \system, chooses to model the pipeline distributed in-network programs using the actor paradigm, with interactions captured through the CSP model.

\begin{figure}[t]
\centering\includegraphics[width=1\linewidth]{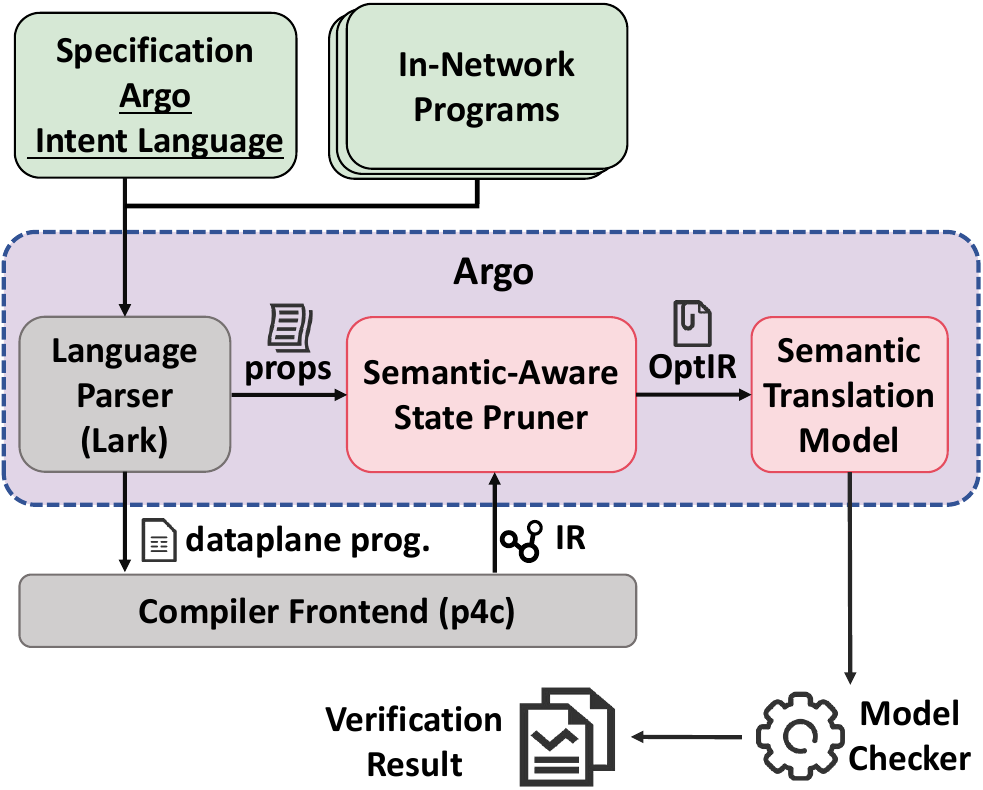}
    \caption{The workflow of \system. \system accepts the specification and in-network programs as input. 
    \texttt{props} denotes the verification intent-related properties, and \texttt{dataplane prog.} are dataplane programs. 
    \texttt{IR} and \texttt{OptIR} are the original and optimized intermediate representations, respectively.
    %The pruner prunes the IR based on props to obtain an intent-related optimized IR (OptIR). The semantic transformation model converts OptIR into a CSP model and ultimately calls the model checker to verify and obtain the verification result.
    }
    
    \label{fig:workflow}
\end{figure}

% In this section, we describe the end to end workflow of \system.
\section{\system Workflow}
As depicted in Figure~\ref{fig:workflow}, \system takes two inputs: a set of in-network programs, as well as a user-written specification based on \system Intent Language. Its workflow is built on three core components that process these inputs to produce a verifiable formal model.

\noindent{\bf Language Parser.} This component serves as the entry point for the user's Specification. It is a lightweight parser based on the Lark~\cite{lark} toolkit. It parses the file to extract the core Verification Intent, which specifies the properties to be checked. This intent provides the precise goal for the subsequent optimization step. 
Meanwhile, it also calls the compilation front end (in our implementation, p4c) to process the dataplane information, providing a basis for subsequent pruning and optimization.
We describe the syntax and semantics of the intent language in detail in \cref{sec:spec}.

\noindent{\bf Semantic-Aware State Pruner.} Concurrently, the in-network programs are processed by the compiler frontend (p4c) to generate an Intermediate Representation (IR). The State Pruner then acts as \system's optimization engine. As shown in Figure~\ref{fig:workflow}, it receives two inputs: the verification intent-related properties from the language parser and the IR from the compiler. Based on this intent, it identifies and eliminates irrelevant program logic, generating a smaller, optimized IR. We detail this pruning algorithm in \cref{sec:pruning}.

\noindent{\bf Semantic Translation Model.} Ultimately, \system transforms the optimized IR into a Semantic Translation Model. The Semantic Translation Model is a two-layer model proposed by \system. It simulates the interaction behaviors between different devices through CSP and the complex semantic behaviors within the devices through actors. We described this model in detail in \cref{sec:formal-model}.

\vspace{-1em}
\section{Semantic Translation Model}
\label{sec:formal-model}

\system introduces a semantic translation model that maps P4 semantics in distributed in-network programs to actor and CSP representations. It formalizes two key aspects to capture the interactive behaviors: (1) the reactive, event-driven nature of the pipeline, modeled as a reactive actor; and (2) diverse packet interaction primitives, unified under a message-passing abstraction. 
Appendix~\Cref{app:pardepar} has more details on the abstraction of parsing and deparsing rules to support the conditional and mirror semantics.
%\vspace{-1em}

\subsection{Modeling the Pipeline as a Reactive Actor}
\label{subsec:reactive}

\system models the pipeline as a \emph{Reactive Actor} driven by two internal channels: an ingress channel $I$ (\texttt{IngressChan}) and an egress channel $E$ (\texttt{EgressChan}). This design captures the event-driven nature of switches responding to packet arrivals and naturally models self-messaging primitives such as \texttt{recirculate}. The actor follows an \emph{Ingress-Triggered, Egress-Draining} model, where each packet arrival initiates a complete reaction: the ingress packet arrival triggers pipeline processing, and any resulting operations are drained from the egress channel before the next reaction begins.

\noindent{\bf Preliminaries.}
The \emph{actor state} is defined as a pair \((S, I)\), where 
\(S\) represents the program-visible dataplane state, including packet headers, metadata, and persistent structures like registers, 
and \(I\) is the FIFO ingress channel (\texttt{IngressChan}).
The egress channel \(E\) is treated as a transient argument to the draining judgment, not as part of the persistent actor state.

The pipeline logic is abstracted by two atomic functions:
\begin{itemize}
    \item \(\Pi_I(S, p) = (S', E_{\mathrm{out}})\): atomic ingress execution on input packet \(p\), returning an updated state \(S'\) and an initial egress channel \(E_{\mathrm{out}}\).
    \item \(\Pi_E(S, I, p) = (S', I', L_p, E_{\text{add}})\): atomic egress execution on packet \(p\). It may update the dataplane state to \(S'\), the ingress channel to \(I'\) (for recirculation), and returns both a list of external events \(L_p\) and a list of packets \(E_{\text{add}}\) to be prepended to the egress channel (for i2e and e2e cloning).
\end{itemize}

\noindent{\bf Channel (FIFO Queue) Operations.}
All channels in the model are FIFO queues. We define the core operations: 
(1) \(\texttt{[]}\): an empty channel;
(2) \(\texttt{enq}(Q, p)\): Add element \(p\) to queue \(Q\);
(3) \(\texttt{deq}(Q)\): dequeueing the head element \(p\) of a non-empty queue \(Q\);
(4) \(\texttt{is\_empty}(Q)\): checking if \(Q\) is empty.

\begin{comment}

\begin{itemize}
    \item \(\texttt{[]}\): An empty channel. 
    \item \(\texttt{enq}(Q, p)\): Add element \(p\) to queue \(Q\) and return \(Q\).
    \item \(\texttt{deq}(Q)\): \textbf{Dequeue}. This operation is only valid for non-empty queues. It returns a tuple \((p, Q')\), where \(p\) is the head of the original queue \(Q\), and \(Q'\) is the remaining queue after the dequeue operation.
    \item \(\texttt{is\_empty}(Q)\): \textbf{Is Empty}. Returns true if \(Q\) contains no elements, and false otherwise.
\end{itemize}

\end{comment}

\noindent{\bf Observable External Events.}
An observable external event is defined as enqueuing a packet into any ingress channel $I$. 
This captures two behaviors: (1) standard forwarding, where the packet from an actor is enqueued to another actor's ingress channel, (2) \texttt{recirculate}, where the packet is enqueued back to the same actor’s ingress channel.
In contrast, an \texttt{enq} operation on an egress channel $E$ is part of the actor's internal processing and does not constitute an observable event.

We denote an external event as a pair \((c, p')\), representing the operation \(\texttt{enq}(c, p')\), where \(c\) is the target ingress channel. The symbol \(L\) represents a finite sequence of such events.

\noindent{\bf Operation Semantics.}
The core of our model is the complete, atomic reaction of an actor to a single packet, defined by Rule~(\ref{eq:actor-react}). This process is triggered when the actor's ingress channel \(I\) is not empty. It uses the \(\texttt{deq}\) operation to atomically retrieve the head packet \(p_{\mathrm{in}}\) and the remaining ingress channel \(I_0\). The packet \(p_{\mathrm{in}}\) is then processed by the ingress pipeline, and the results are fully drained from the egress channel, producing a sequence of external events \(L\). Finally, the actor's state is updated with the new ingress channel \(I_0\).

{\small
\begin{equation} \label{eq:actor-react}
\frac{
    (p_{\mathrm{in}}, I_0) = \texttt{deq}(I)
    \quad
    \langle S_0, p_{\mathrm{in}} \rangle \xrightarrow{\Pi_I} \langle S_1, E_1 \rangle
    \quad
    \langle S_1, E_1 \rangle \xrightarrow{\text{Drain}} \langle S_2, L \rangle
}{
    \langle S_0, I \rangle \xrightarrow{L} \langle S_2, I_0 \rangle
}
\end{equation}
}

The rule~(\ref{eq:actor-react}) relies on a key helper process, \texttt{Drain}, which recursively processes all packets in the egress channel. This process is defined by the following two rules. Rule~(\ref{eq:drain-empty}) is the base case: if the channel is empty, the process terminates.

{\small
\begin{equation} \label{eq:drain-empty}
\frac{
    \texttt{is\_empty}(E) = \text{true}
}{
    \langle S, I, E \rangle \xrightarrow{\text{Drain}} \langle S, I, [] \rangle
}
\end{equation}
}

Rule~(\ref{eq:drain-step}) is the recursive step, now enhanced to support packet re-injection. It dequeues a packet \(p_e\), processes it via \(\Pi_E\), and obtains a list of packets to re-inject, \(E_{\text{add}}\). These packets are prepended to the remaining queue \(E_0\) to form the new queue \(E'\) for the recursive call.

{\small
\begin{equation} \label{eq:drain-step}
\frac{
    \begin{gathered}
    (p_e, E_0) = \texttt{deq}(E)
    \quad
    \langle S_0, I_0, p_e \rangle \xrightarrow{\Pi_E} \langle S_1, I_1, L_p, E_{\text{add}} \rangle \\
    E' = \texttt{concat}(E_{\text{add}}, E_0)
    \quad
    \langle S_1, I_1, E' \rangle \xrightarrow{\text{Drain}} \langle S_2, I_2, L_0 \rangle
    \end{gathered}
}{
    \langle S_0, I_0, E \rangle \xrightarrow{\text{Drain}} \langle S_2, I_2, L_p \cdot L_0 \rangle
}
\end{equation}
}

This model effectively divides packet processing into two phases. The ingress pipeline ($\Pi_I$) acts as an event handler, consuming one packet via the top-level \texttt{Actor-React} rule (Rule~\ref{eq:actor-react}). The egress pipeline ($\Pi_E$), invoked via the \texttt{Drain} process (Rules~\ref{eq:drain-empty} and~\ref{eq:drain-step}), then executes all resulting actions to produce the final, observable external events. The specific behaviors of $\Pi_I$ and $\Pi_E$ are determined by the semantics of the \texttt{TrafficManager} and \texttt{FinalDispatch} stages, which we formalize in the next section \Cref{subsec:messagepass}.

\subsection{Unifying Interaction as Message Passing}
\label{subsec:messagepass}

We unify all interactive behaviors that move a packet across logical boundaries, such as forwarding, cloning, and recirculation, and abstract them as message passing between actors. This abstraction is captured by the semantics of two architectural stages: \texttt{TrafficManager} (TM) and \texttt{FinalDispatch} (FD). These stages interpret packet metadata to resolve the programmer’s intent, deciding whether to deliver the message to an external actor (forwarding) or back to the same actor (recirculation). 

\subsubsection{TrafficManager (TM) Semantics}
The \texttt{TrafficManager} is the final logical stage of the ingress pipeline $\Pi_I$. Its responsibility is to populate the egress channel ($E$) based on the packet's metadata, particularly \texttt{clone\_spec}. Its behavior is defined as a function $\text{TM}(p) = E_{\text{out}}$. We also define the helper functions for cloning:
\begin{itemize}
    \item \(\texttt{clone}(p)\): This function creates an exact copy of a packet, \(p_{\text{clone}}\).
    \item \(\texttt{clone\_and\_mark\_i2i}(p)\): This function performs the \(\texttt{clone}(p)\) operation and additionally sets a metadata flag on the clone: \(p'_{\text{clone}}.\text{meta.is\_i2i\_clone} \leftarrow \text{true}\).
\end{itemize}

Rule~(\ref{eq:tm-normal}) defines the default case. When no special cloning is required, the TM creates a new egress channel containing only the current packet.

{\small
\begin{equation} \label{eq:tm-normal}
\frac{
    p.\text{meta.clone\_spec} = \text{None}
}{
    \text{TM}(p) = \texttt{enq}([], p)
}
\end{equation}
}

The rules for other cloning types (e.g., ingress-to-egress) and packet drops within the ingress pipeline are detailed in Appendix~\cref{app:tmse}.

\subsubsection{FinalDispatch (FD) Semantics}
\label{subsubsec:fnldpc}

The \texttt{FinalDispatch} is the final logical stage of the egress pipeline $\Pi_E$ and the sole exit point for all packets. Its responsibility is to translate a packet \(p\) being processed in the egress pipeline into a tuple \((L_p, E_{\text{add}})\), where \(L_p\) is a list of formal representations of external \(\texttt{enq}\) events, and \(E_{\text{add}}\) is a list of packets to be re-injected into the egress pipeline.

The behavior of \texttt{FD}, particularly for forwarding, is critically dependent on the network topology, which we abstract as a \texttt{Links} relation. This relation is a set of tuples of the form \((\text{Source Actor, Egress Port, Destination Channel})\). If a packet's destination does not match any of the following rules, it is implicitly dropped (i.e., \(\text{FD}(p) = ([], [])\)), with further details in Appendix~\cref{app:fdse}.

Rule~(\ref{eq:fd-forward}) defines standard, topology-aware forwarding. An external event is generated only if a valid link exists in the \texttt{Links} relation.

{\small
\begin{equation} \label{eq:fd-forward}
\frac{
    \begin{gathered}
    p.\text{meta.recirculate\_flag} = \text{false} \\
    (X, p.\text{meta.egress\_port}, c_{\text{dest}}) \in \text{Links}
    \end{gathered}
}{
    \text{FD}(p) = ([ (c_{\text{dest}}, p') ], [])
}
\end{equation}
}

Rule~(\ref{eq:fd-recirculate}) models recirculation. This is an internal behavior that generates an event targeting the actor's own ingress channel, \(I_{\text{self}}\).
{\small
\begin{equation} \label{eq:fd-recirculate}
\frac{
    p.\text{meta.recirculate\_flag} = \text{true}
}{
    \text{FD}(p) = ([ (I_{\text{self}}, p') ], [])
}
\end{equation}
}

Rule~(\ref{eq:fd-e2e-clone}) models egress-to-egress (E2E) cloning. It produces one external event for the original packet and re-injects the clone packet into the egress pipeline, allowing the clone to have its own lifecycle.
{\small
\begin{equation} \label{eq:fd-e2e-clone}
\frac{
    \begin{gathered}
    p.\text{meta.clone\_spec} = \text{E2E} \quad p_{\text{clone}} = \texttt{clone}(p) \\
    (X, p.\text{meta.egress\_port}, c_{\text{dest}}) \in \text{Links}
    \end{gathered}
}{
    \text{FD}(p) = ([ (c_{\text{dest}}, p') ], [ p_{\text{clone}} ])
}
\end{equation}
}

The rule models for the ingress-to-ingress mirroring, and packet drops within the egress pipeline are detailed in Appendix~\ref{app:fdse}.

\noindent\textbf{Model Boundary Discussion.} It is worth noting that since \(\texttt{clone}(p)\) is an exact copy, our model allows a cloned packet to be cloned again, which could theoretically lead to non-termination. However, in practical network applications, protocol designs that require infinite mirroring are exceedingly rare. We therefore do not explicitly model a termination condition in order to maintain the simplicity of the core model.

\vspace{-1em}
\subsection{Executable Translation for Promela}
\label{app:translation-executable}

To automate the verification of distributed properties, we translate our formal operational semantics into an executable model in Promela, the specification language of SPIN~\cite{spin}. Promela's \texttt{proctype} and \texttt{chan} directly instantiate our global actor model: devices become concurrent processes; inter-device links are FIFO channels. 
We now give a translation that preserves the reaction-level semantics (\cref{subsec:reactive}) and the observable interface (\emph{enqueue into any ingress}).

\noindent{\bf System composition in Promela.}
The parallel composition $\mathrm{System} = D_1 \parallel \dots \parallel D_n$ becomes the \texttt{init} block. We declare one \emph{observable} ingress channel per device and spawn a \texttt{proctype} per device, passing its ingress handle.
{%
\fontsize{10pt}{12pt}\selectfont
\begin{verbatim}
init {
    chan s1_ig_chan = [Q_IN] of { packet_t };
    chan s2_ig_chan = [Q_IN] of { packet_t };

    run device_proc(1, s1_ingress);
    run device_proc(2, s2_ingress);
}
\end{verbatim}
}

\noindent{\bf Device pipeline as a \textit{proctype}.}
Each device implements the Reactor pattern and executes a whole \emph{reaction} inside one \texttt{atomic\{\}}: dequeue one ingress packet, run $\Pi_I$, populate the local egress queue via TM, then \emph{drain} the egress queue by repeatedly running $\Pi_E$ and dispatching via FD. The reaction ends only after the egress queue becomes empty, matching Rule (\ref{eq:drain-empty}) and (\ref{eq:drain-step}).
{%
\fontsize{10pt}{12pt}\selectfont
\begin{verbatim}
proctype device_proc()
{
    chan egress_q = [Q_EG] of { packet_t }; 
    packet_t p; 
    packet_t pe;
    do
    :: ingress_chan ? p ->
       atomic {  
           IngressParser(p);    
           Ingress(p);         
           IngressDeparser(p);
           TrafficManager(p);  
           do
           :: len(egress_q) > 0 ->
                egress_q ? pe;
                EgressParser(pe);
                Egress(pe);         
                EgressDeparser(pe);
                FinalDispatch(self, 
                my_ingress, pe);
           :: else -> break
           od;
       }
    od
}
\end{verbatim}
}

\noindent{\bf Traffic Manager inline function.}
To realize Rule (~\ref{eq:actor-react}), we materialize \(\Pi_I\)’s \(E_{\mathrm{out}}\) as pure \emph{local} appends to the device egress queue: `drop` elides the packet, `I2E/I2I` duplicate locally and append, and no channel send occurs; thus TM yields no observables within one reaction.
{%
\fontsize{10pt}{12pt}\selectfont

\vspace{10pt}

\begin{verbatim}
inline TrafficManager(packet_t p)
{
    if
    :: p.drop_flag ->
         skip                      
    :: else ->
         if
         :: p.clone_spec == I2E ->
              packet_t c = p;
              c.egress_spec = get_mirror_port(p);
              egress_q ! p;         
              egress_q ! c;   
         :: p.clone_spec == I2I ->
              packet_t c2 = p;
              c2.is_i2i_clone = true;  
              egress_q ! p;
              egress_q ! c2;
         :: else ->
              egress_q ! p
         fi
    fi
}
\end{verbatim}
}

\noindent{\bf Final Dispatch inline function.}
Consistent with Rules~(\ref{eq:fd-forward}, \ref{eq:fd-recirculate}, \ref{eq:fd-e2e-clone}), FD deterministically maps flags to a single observable action: \texttt{recirc}/\texttt{is\_i2i\_clone} (with capacity \texttt{assert}s) self-enqueue to \texttt{my\_ingress}; otherwise it forwards via \texttt{egress\_spec}. FD is the sole boundary where internal effects become channel sends. See Appendix~\ref{app:translation-appendix} for \texttt{send\_to()}.
{%
\fontsize{10pt}{12pt}\selectfont
\begin{verbatim}
inline FinalDispatch(packet_t pe)
{
    if
    :: pe.drop_flag ->
         skip     
    :: pe.recirc_flag ->
         assert(len(my_ingress) < Q_IN);
         my_ingress ! pe     
    :: pe.is_i2i_clone ->
         assert(len(my_ingress) < Q_IN);
         my_ingress ! pe      
    :: else ->
         send_to(self, pe.egress_spec, pe)
    fi
}
\end{verbatim}
}

\vspace{-1em}
\section{Semantic-Aware State Pruner}\label{sec:pruning}

\begin{algorithm}[t]
\caption{Semantic-Aware DDG Construction from CFG}
\label{alg:ddg}
\begin{algorithmic}[1]
\Require $CFG=(N,E,\textit{entry},\textit{exit})$
\Ensure $DDG=(N, E_D)$
\Function{BuildDDG}{$CFG$}
    \State $G \gets$ empty directed graph
    \State $RD_1 \gets \Call{ReachingDefs}{CFG}$ \label{alg:l:phase1-rd}
    \ForAll{$u \in N$}                                   % base pass
        \ForAll{$x \in \Call{Uses}{u}$}
            \State $D_1[u,x] \gets \{\, d \in N \mid \Call{Defines}{d,x} \wedge d \in RD_1[u] \,\}$
            \ForAll{$d \in D_1[u,x]$}
                \State $\Call{AddEdge}{G, d, u, x}$ \label{alg:l:phase1-gen}
            \EndFor
        \EndFor
    \EndFor
    \If{$\Call{NeedsLoopEdge}{CFG}$} \label{alg:l:phase2-check}
        \State $CFG^+ \gets (N,\, E \cup \{(\textit{exit},\textit{entry})\},\, \textit{entry},\, \textit{exit})$ \label{alg:l:add-back-edge}
        \State $RD_2 \gets \Call{ReachingDefs}{CFG^+}$ \label{alg:l:phase2-rd}
        \ForAll{$u \in N$}                               % loop-aware pass
            \ForAll{$x \in \Call{Uses}{u}$}
                \State $D_2 \gets \{\, d \in N \mid \Call{Defines}{d,x} \wedge d \in RD_2[u] \,\}$
                \State $New \gets D_2 \setminus D_1[u,x]$ \label{alg:l:new-only}
                \ForAll{$d \in New$}
                    \If{$\Call{CrossOK}{x}$}             % boolean guard for cross-iteration carry
                        \State $\Call{AddEdge}{G, d, u, x}$ \label{alg:l:crossok-edge}
                    \EndIf
                \EndFor
            \EndFor
        \EndFor
    \EndIf
    \State \Return $G$
\EndFunction

\Statex

\Function{CrossOK}{$x$} 
    \State \Return $\Call{IsRegister}{x}\ \vee\ \Call{IsPacketCarried}{x}$
\EndFunction
\end{algorithmic}
\end{algorithm}

A key challenge in verification is state-space explosion, where the number of possible states grows exponentially with concurrent actors and their interactions. Directly executing our actor-based model in a model checker can generate an enormous state space, resulting in high computational overhead.

To tackle this, we propose a semantic-aware state pruner, which aims to simplify the intermediate representation (IR) and reduce model-checking time. The core idea is to eliminate states that do not affect the properties being verified.

However, traditional static analysis tools, which operate on static Control Flow Graphs, treat the P4 pipeline as a single-pass execution graph. This view cannot capture primitives like recirculation, where a packet exits the pipeline only to be reinjected for a subsequent pass. The root cause of this issue is the failure to recognize the ``cross-pass data dependency'': the state modified in one pass becomes a critical input that governs the control flow of the next pass.

\noindent\textbf{Dependency Graph Construction.}
Specifically, we propose a semantic-aware data dependency graph construction algorithm (Algorithm~\ref{alg:ddg}) that takes only the program’s control flow graph (CFG) as input and produces a data dependency graph (DDG). The algorithm proceeds in two passes. In the first pass, it runs a classical \emph{reaching-definitions} analysis on the original CFG (line~\ref{alg:l:phase1-rd}) and creates a baseline DDG by adding one edge $d \to u$ labeled with variable $x$ whenever $x$ is used at node $u$ and a definition of $x$ at node $d$ reaches $u$ (line~\ref{alg:l:phase1-gen}). If the program contains a pipeline loop (\eg due to \texttt{recirculate}/\texttt{mirror}), the second pass materializes it by inserting a \emph{single} back edge from \textit{exit} to \textit{entry} (line~\ref{alg:l:add-back-edge}), recomputes reaching definitions on the augmented CFG (line~\ref{alg:l:phase2-rd}), and then adds only those \emph{new} cross-iteration dependency edges that did not exist after the first pass (line~\ref{alg:l:new-only}). Each such edge is guarded by a boolean predicate \textsc{CrossOK}$(x)$ (line~\ref{alg:l:crossok-edge}) that authorizes cross-iteration carry of $x$ (true for persistent state, such as registers, or for packet-carried fields).

Algorithm~\ref{alg:prune} performs a backward slicing over the union of dependencies:
starting from the property-related seeds, it traverses all reachable statement nodes via
data edges and control edges.
All nodes that are backward-reachable constitute $Keep$ and are retained; others can be pruned.

\begin{figure}[t]
\centering\includegraphics[width=0.99\linewidth]{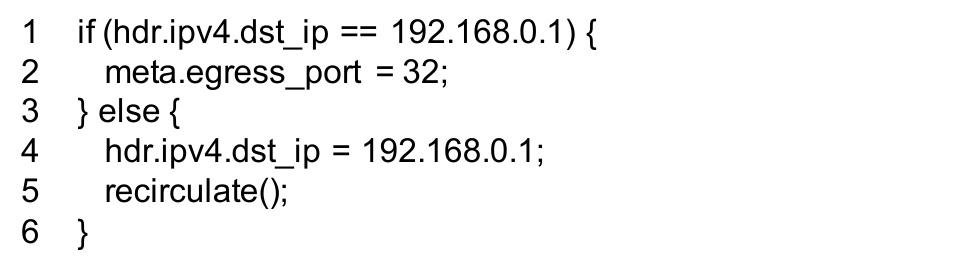}
    \caption{An example of a P4 backward dependency.}
    \label{fig:prun_code}
\end{figure}

\begin{figure}
\centering
\begin{subfigure}{0.32\linewidth} 
    \centering
    \includegraphics[width=0.99\linewidth]{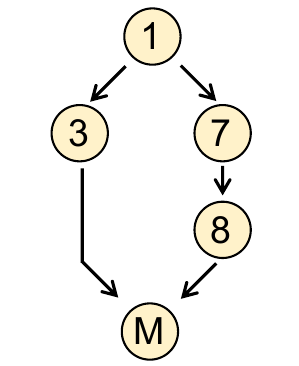}
    \caption{Original CFG.}
    \label{fig:prun_cfg_a}
\end{subfigure}
\begin{subfigure}{0.32\linewidth} 
    \centering
    \includegraphics[width=0.99\linewidth]{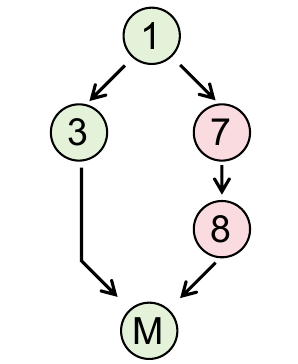}
    \caption{Error pruning.}
    \label{fig:prun_cfg_b}
\end{subfigure}
\begin{subfigure}{0.32\linewidth} 
    \centering
    \includegraphics[width=0.99\linewidth]{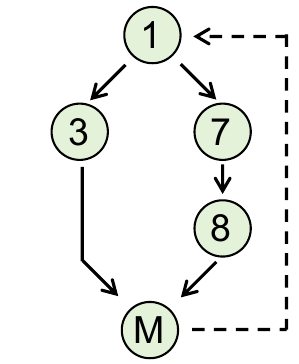}
    \caption{Correct Slicing.}
    \label{fig:prun_cfg_c}
\end{subfigure}
\caption{CFG of example in Figure~\ref{fig:prun_code}.}
\vspace{-5mm}
\label{fig:prun_cfg}
\end{figure}

In Figure~\ref{fig:prun_code}, the code exhibits a pipeline loop (\eg a \texttt{recirculate}) where a field updated in one pass
decides a branch and triggers an action in the next pass. When the DDG already encodes this with the single
\emph{exit$\rightarrow$entry} back edge from the previous section, the backward traversal naturally climbs from the seed
location in the second pass to the writing site in the first pass via DDG, while CDG preserves the necessary branch
conditions along the way. By contrast, Figure~\ref{fig:prun_cfg_b} illustrates the failure of the single-pass dependency
view: missing the cross-iteration link, the traversal cannot reach the prior definition and wrongly drops key statements.
Our approach (shown in ~\ref{fig:prun_cfg_c}) recovers the cross-iteration chain and keeps exactly those statements.

% Specifically, we propose a semantic-aware data dependency graph construction algorithm (Algorithm~\ref{alg:dep-construction}) that takes the program's control flow graph (CFG) and its original IR as input. For example, the CFG of the code in Figure~\ref{fig:prun_code} is shown in Figure~\ref{fig:prun_cfg_a}, with line numbers as node IDs.
% The algorithm first generates a data dependency graph using classical reaching definitions analysis (\texttt{COMPUTEREACHDEFS()}) on the original CFG  (line~\ref{alg:line:phase1-gen}). If the program contains recirculate or mirror operations, the corresponding edges are added to the CFG (\eg Figure~\ref{fig:prun_cfg_c} adds an edge to Figure~\ref{fig:prun_cfg_a} as a recirculate exists in Figure~\ref{fig:prun_code}). A second reaching-definitions analysis is then performed on the updated CFG to produce an enhanced data dependency graph (lines~\ref{alg:line:phase2-check}-\ref{alg:line:phase2-gen}).
% Finally, the algorithm traverses both dependency graphs to determine whether each node is relevant to the verification properties, pruning nodes with no dependencies (lines~\ref{alg:aaa}-\ref{alg:bbb}). This effectively reduces the state space in the IR.

% \qs{introduce the details of the algorithm DataEdgesGen.}

\noindent{\bf Analysis.}
After pruning, the state in IR retains only the minimal set of packet headers, metadata fields, and registers that can influence property verification. Likewise, the CFG preserves only conditional branches that affect the execution of statements deemed essential in the data dependency graph. Execution paths irrelevant to these state variables are eliminated. As a result, both the state vector size and the overall state space are significantly reduced, alleviating the computational burden of subsequent model checking.

\begin{algorithm}[t]
\caption{Backward Slicing-based Pruning}
\label{alg:prune}
\begin{algorithmic}[1]
\Require Data-dependency graph $DDG=(V,E_D)$; Control-dependency graph $CDG=(V,E_C)$; seed nodes $Seeds \subseteq V$
\Ensure Set of kept nodes $Keep \subseteq V$
\Procedure{BackwardPrune}{$DDG, CDG, Seeds$}
    \State $Keep \gets \emptyset$;\quad $Work \gets Seeds$
    \While{$Work \neq \emptyset$}
        \State $u \gets \Call{Pop}{Work}$
        \If{$u \in Keep$} \State \textbf{continue} \EndIf
        \State $Keep \gets Keep \cup \{u\}$
        \ForAll{$(p \rightarrow u) \in E_D$} \State $\Call{Push}{Work, p}$ \EndFor
        \ForAll{$(c \rightarrow u) \in E_C$} \State $\Call{Push}{Work, c}$ \EndFor
    \EndWhile
    \State \Return $Keep$
\EndProcedure
\end{algorithmic}
\end{algorithm}

\section{\system Intent Language}
\label{sec:spec}

We now introduce the \system Intent Language for specifying verification properties of distributed in-network programs with complex interactive behaviors. It unifies three essential components: (1) topology behavior specification, defining switches and their interconnection topology; (2) local behavior specification, specifying contracts and assertions in a single switch; and (3) global behavior specification, expressing temporal properties across multiple switches.

\begin{comment}

\noindent{\bf System Activation and Host Modeling.}
\textsc{Procurator} models the P4 network as a passive, reactive Actor system that requires external stimulus to be activated. Therefore, the P4 programs and topology alone are insufficient to form a complete, verifiable model; we must also define the source of this stimulus---the behavior of a \textbf{Host}. The Intent Language addresses this through a flexible workflow:

First, for a given P4 system, an engineer can use \textsc{Procurator} to analyze the programs and generate a global message definition file. Based on this file, the engineer can manually write a Host model to simulate the behavior of external servers or clients.

Subsequently, using the \texttt{import} statements within the intent language, \textsc{Procurator} compiles and links the specified P4 programs, the topology, and the user-written Host model into a complete, executable verification suite, upon which model checking is performed.

\end{comment}

%\subsection{Core Syntax}

A \system intent specification consists of a (topology, local intent, global intent) triple. 
Figure~\ref{fig:grammar} presents the syntax.

\begin{figure}[t]
\centering
% The following command adjusts the vertical spacing between lines in the table for better readability.
\renewcommand{\arraystretch}{1.2} 
% The tabular environment uses a specific format:
% r @{...} l defines two columns: a right-aligned column for the non-terminal,
% and a left-aligned column for the production rule, separated by a nicely spaced ::=
\begin{tabular}{r @{\ \ $::=$\ \ } l}
    % Top-level rules for the specification and declarations
    \textit{spec} & $\textit{decl}^{+}$ \\
    \textit{decl} & \textbf{import} \textit{name} \textbf{from} \textit{string} [ \textbf{entries} \textit{string} ] \textbf{;} \\
                 & $|$ \textit{topo} \\
                 & $|$ \textit{local} \\
                 & $|$ \textit{global} \\[1ex] % Adds a little extra vertical space after the group

    % Rules for topology definition
    \textit{topo} & \textbf{topology} { \textit{link}* } \\
    \textit{link} & \textbf{link} \textit{dev} -> \textit{dev} (\textit{port\_id} $|$ \textbf{ALL}) \textbf{;} \\[1ex]

    % Rules for local, device-specific intents
    \textit{local} & \textbf{local} \textit{dev} { \textit{local\_stmt}* } \\
    \textit{local\_stmt} & \textbf{let} \textit{id} = \textit{expr} \textbf{;} \\
                 & $|$ \textbf{assert} \textit{expr} \textbf{;} \\[1ex]

    % Rules for global, system-wide intents using LTL
    \textit{global} & \textbf{global} { \textit{ltl\_rule}* } \\
    \textit{ltl\_rule} & \textbf{ltl} \textit{name} { \textit{ltl\_expr} } \textbf{;} \\
    \textit{ltl\_expr} & \textit{temporal\_op} { \textit{expr} } \\
                 & $|$ \textit{ltl\_expr} \textit{ bin\_op } \textit{ltl\_expr} \\
                 & $|$ \textit{un\_op} \textit{ltl\_expr} \\
                 & $|$ ( \textit{ltl\_expr} ) \\[1ex]

    % Definition of LTL temporal operators
    \textit{temporal\_op} & `[]' $|$ `<>' $|$ `X' $|$ `U' \\[1ex]

    % Definition of expressions used in assertions and LTL
    \textit{expr} & \textit{id} $|$ \textit{literal} \\
                 & $|$ \textit{expr} \textit{ arith\_op } \textit{expr} \\
                 & $|$ \textit{expr} \textit{ logic\_op } \textit{expr} \\
                 & $|$ \textit{id} . \textit{field} \\
\end{tabular}
\caption{The abstract syntax of the \system Intent Language.}
\vspace{-5mm}
\label{fig:grammar}
\end{figure}

\begin{figure}[t]
\centering\includegraphics[width=0.99\linewidth]{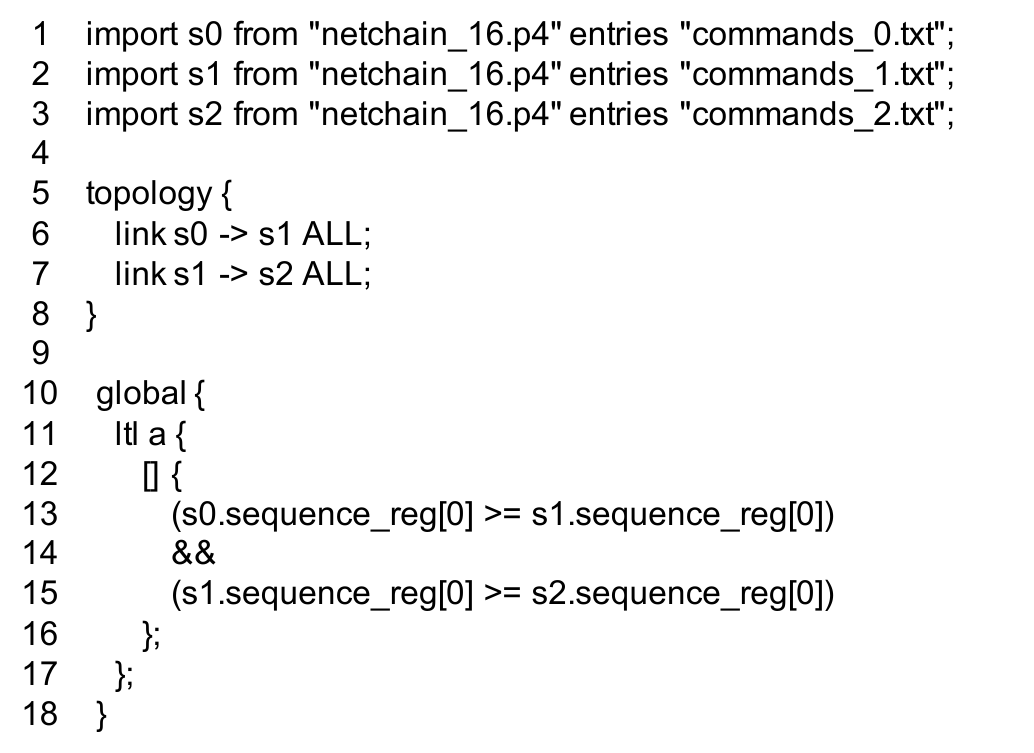}
    \caption{
Specification S of NetChain in~\system.}
    \label{fig:netchain_spec}
\end{figure}

\noindent{\bf Topology behavior specification.}
As shown in the syntax for \textit{topo} and \textit{link}, the \texttt{topology} block represents the network as a directed graph. Each \texttt{link} statement is closely tied to the \texttt{FinalDispatch} stage (\Cref{subsubsec:fnldpc}). The declaration \texttt{link X -> Y port\_id} provides routing information to the code generator: when switch X’s actor forwards a packet from logical port \textit{port\_id}, the generator translates it into a message sent to switch Y’s \texttt{IngressChannel} using Rule(\ref{eq:fd-forward}). The syntax also supports the \texttt{ALL} keyword, which specifies a default, unconditional forwarding path.

\noindent{\bf Local behavior specification.}
The \texttt{local} block enables engineers to specify functional contracts for the program in a single switch. According to the \texttt{local\_stmt} rule in Figure~\ref{fig:grammar}, it supports two statement types:

\begin{itemize}
    \item \textit{Variable Declaration and Computation}: Using the \texttt{let} keyword, engineers can declare temporary variables and assign them the result of \textit{expr}. This expression can involve arithmetic and logical operations on state variables like headers or registers.
    \item \textit{Assertion}: Using the \texttt{assert} keyword, engineers can check whether a boolean \textit{expr} holds true.
\end{itemize}

Assertions are automatically injected just before the \texttt{FinalDispatch} stage, allowing precise checks on the pipeline state before generating external events. Variables declared with \texttt{let} are scoped to the local block but can also be referenced in the \texttt{global} block (for global behavior specification), enabling verification logic that spans both local computations and global properties. During parsing, \system performs semantic checks to ensure all variables originate from the actor's state or are locally declared.

\noindent{\bf Global behavior specification.}
The \texttt{global} block specifies system-wide properties using Linear Temporal Logic (LTL). The syntax for \textit{ltl\_rule} and \textit{ltl\_expr} defines named LTL properties, supporting standard temporal operators such as \texttt{[]} (always), \texttt{<>} (eventually), \texttt{X} (next), and \texttt{U} (until).

LTL expressions can reference state variables from any actor (\eg headers, metadata, or registers) as well as variables declared in \texttt{local} blocks. During parsing, \system performs semantic checks to ensure all variables are valid, preventing specification errors.

\noindent{\bf A concrete example.}
We now provide a concrete example to verify the sequence monotonicity property in NetChain.
As shown in Figure~\ref{fig:netchain_spec}, the \texttt{topology} block explicitly defines the linear structure \texttt{s0->s1->s2}. The \texttt{global} block specifies a single LTL property, \texttt{sequence\_monotonicity}, which uses the always operator (\texttt{[]}) to assert an invariant: at all times, the \texttt{sequence\_reg} value in an upstream switch must be greater than or equal to that in its downstream neighbor. This cross-actor state assertion captures a distributed property that traditional verifiers for the program on an individual switch cannot express.

\begin{table*}[ht]
\centering
\resizebox{\textwidth}{!}{
\begin{tabular}{|c|c|l|c|c|c|}
\hline
\textbf{System} & \textbf{Property} & \textbf{Root Cause} & \textbf{States} & \textbf{Time} & \textbf{LoC of Host and Spec}\\
\hline
\multirow{1}{*}{\textbf{Netchain}~\cite{netchain}} 
& Sequence monotonicity 
& Sequence number wraparound 
& 229374 
& 1min39s
& 162 \\
\hline

\multirow{3}{*}{\textbf{P4xos}~\cite{xos}} 
& Majority quorum 
& Incorrect quorum computation due to persistent stale counts
& 10765 
& 1min27s
& 141 \\
& Register access safety 
& Out-of-bound register access* 
& 4507 
& 35.3s
& 108 \\
& Forwarding correctness 
& \texttt{Drop\_flag} not being correctly set* 
& 53 
& 0.45s
& 93 \\
\hline

\multirow{3}{*}{\textbf{Gecko}~\cite{gecko}} 
& Timer Packet Loss or Blockage 
& Single timer packet is lost, halting timeout updates 
& 8526 
& 41.6s
& 515 \\
& Limited Concurrency Handling 
& Timer logic can only process one request’s per traversal 
& 1348 
& 4.68s
& 516 \\
& Improper Timer Initialization 
& Early write requests, bypassing in-switch logic 
& 2679 
& 9.17s
& 520 \\
\hline

\multirow{1}{*}{\textbf{DistCache}~\cite{distcache}} 
& Integer Wrap-around in Load Tracking 
& \texttt{leafload} variable overflows, causing misrouting
& 163858 
& 18.5s
& 162 \\
\hline

\multirow{2}{*}{\textbf{ATP}~\cite{atp}} 
& Indefinite Aggregation 
& 8-bit \texttt{agtr\_time} set above intended range 
& 72 
& 0.17s
& 129 \\
& Aggregation Count Mismatch 
& Packet loss leads to fewer completed 
& 390023 
& 5min30s
& 148 \\
\hline
\end{tabular}
}
\caption{Effectiveness and efficiency in detecting bugs. \textnormal{The "LoC of Host" column indicates the lines of Promela host-model code required for each system. 
Results shown here assume no additional optimizations in place. 
(* indicates old bugs previously found by existing P4 verifiers.)
}}
\vspace{-5mm}
\label{tab:bugs}
\end{table*}

\section{Implementation}
\label{sec:impl}

We implement a prototype of \system~using $\sim$12k lines of C++ and $\sim$2k lines of Python. Our prototype includes a language parser for our unified intent language, a semantic translation model that generates verifiable outputs by the model checker, and a semantic-aware state pruner for optimization.

\noindent{\bf \system Intent Language Parser.} To handle user-defined verification tasks, we implemented an interpreter for \system~Intent Language, leveraging the Lark~\cite{lark} toolkit. This component is responsible for processing a specification file that defines the system's topology, local invariants, and global properties. The output of this parser is a structured representation of the verification intent, which guides both the state pruner and the final model generation.

\noindent{\bf Semantic Translation to Promela.} The semantic translation model converts the in-network program's intermediate representation (IR) into a verifiable model for the SPIN\cite{spin} model checker. We implemented this translator as a custom backend for the p4c\cite{p4c-compiler} in C++. The translator methodically maps P4's operational semantics to the concepts of Communicating Sequential Processes (CSP), which are natively supported by Promela~\cite{spin}. 

\noindent{\bf Semantic-Aware State Pruner.} Our implementation of the pruner functions as an optimization pass that simplifies the P4 compiler's IR before it is fed to the Promela translator. The pruner implements the backward slicing algorithm detailed in \Cref{sec:pruning}, which constructs a precise data dependency graph based on reachingdefinitions analysis.

\begin{comment}
We implement a prototype of \system in around 12k lines of C++ and 2k lines of Python, supporting both V1Model (software switches) and TNA (Tofino Native Architecture) pipelines. It uses \texttt{bf-p4c} or \texttt{p4c} as the front-end compiler and performs a \emph{slicing} optimization on the generated intermediate representation. The sliced program is then translated into a Promela model, which SPIN explores to analyze potential execution paths.

\para{Spec Parser and Translation Flow.}
We employ a \textsc{lark}-based specification parser and interpreter to read user-defined specifications (e.g., the control-plane table entries or desired properties). Once the specification is parsed, \system invokes our customized \emph{translater} (built on \texttt{P4C}) using the compiled IR from the P4 files. We also merge a global \texttt{header} structure, which acts as a \emph{common header} shared across the compiled P4 programs in a multi-device scenario.

\end{comment}

%\vspace{-1em}
\section{Evaluation}\label{sec:evaluation}
\vspace{-1em}
We now present the evaluation of \system by answering the following questions.
All experiments have been performed on a server with 2 Intel Xeon Silver 4210R CPUs and 128GB of memory. 

\begin{itemize}
    \item Can \system effectively detect subtle bugs in existing distributed in-network programs? (\cref{sec:exp-bugs})
    \item Does \system offer advantages over state-of-the-art verifiers for individual P4 programs? (\cref{sec:comparison})
    \item What about the efficiency of the pruning optimization in \system? (\cref{sec:exp-slicing})
\end{itemize}

%We used \system~to perform experiments concerning the following questions: (1) How effective is \system~in detecting subtle bugs of distributed P4 systems? (2) How much efficiency improvement does our State Pruner provide? (3) How does \system~perform against state-of-the-art single-device P4 verifiers?

%\subsection{Setup}\label{sec:exp-setup}

We evaluate \system using two sets of practical P4 programs. 
The first set consists of six real-world stateful distributed in-network systems (Table~\ref{tab:bugs}), 
which exhibit common interaction patterns in practice. This set is used to assess \system in \cref{sec:exp-bugs} and \cref{sec:exp-slicing}. 
The second set includes four individual P4 programs (Table~\ref{tab:comparison}), used for direct comparison with state-of-the-art stateful verifiers for individual programs (\cref{sec:comparison}).

\begin{comment}

\noindent\textbf{Metrics.} 
We evaluate \system~based on two primary categories of metrics: effectiveness and efficiency. 
Effectiveness means the ability of \system~to correctly identify correctness violations. 
We assess whether it can find known bugs uncovered by existing P4 verifiers and, more importantly, discover new, 
subtle bugs in both distributed and complex single-device program that are missed by other tools. 
We measure efficiency using three quantitative metrics: verification time, memory Consumption, and the number of distinct states explored by the model checker SPIN.
    
\end{comment}

\vspace{-1em}
\subsection{Verification on Distributed Programs}
\label{sec:exp-bugs}

%Additionally, the components of these systems, once compiled into Promela, vary in size from 470 to 1600 lines of model, 
%reflecting the diverse complexity of the programs under analysis.

%\qs{todo: the bugs are not found manually.}
%To evaluate Procurator’s ability to uncover bugs in complex real-world systems, we use a verification pipeline below. For each system, we first study its design goals and open-source implementation to distill the core correctness properties the system must satisfy as a specification, then implement a “host model” that encodes a representative set of interaction scenarios we believe are most likely to expose faults. Then, Procurator systematically explores all reachable states to determine whether any violation of the core properties occurs.

We evaluate \system's effectiveness by detecting bugs in five representative systems (Table~\ref{tab:bugs}). 
Since NetChain's bug is already discussed in \Cref{subsub:switch_inter}, we focus here on the other four systems. 
As shown in Table~\ref{tab:bugs}, \system uncovered 10 bugs in total. 
%For SwitchV2P, where no bug was found, we instead provide a case study on scenarios with no property violation.
All detected bugs were confirmed through error-trace analysis. 
Notably, most of these bugs are subtle, deeply embedded, and tied to complex interleavings 
--- cases typically beyond the reach of conventional testing or existing P4 verifiers.

\label{sec:p4xos-result}

\begin{figure}[t] \centering
    \includegraphics[width=\linewidth]{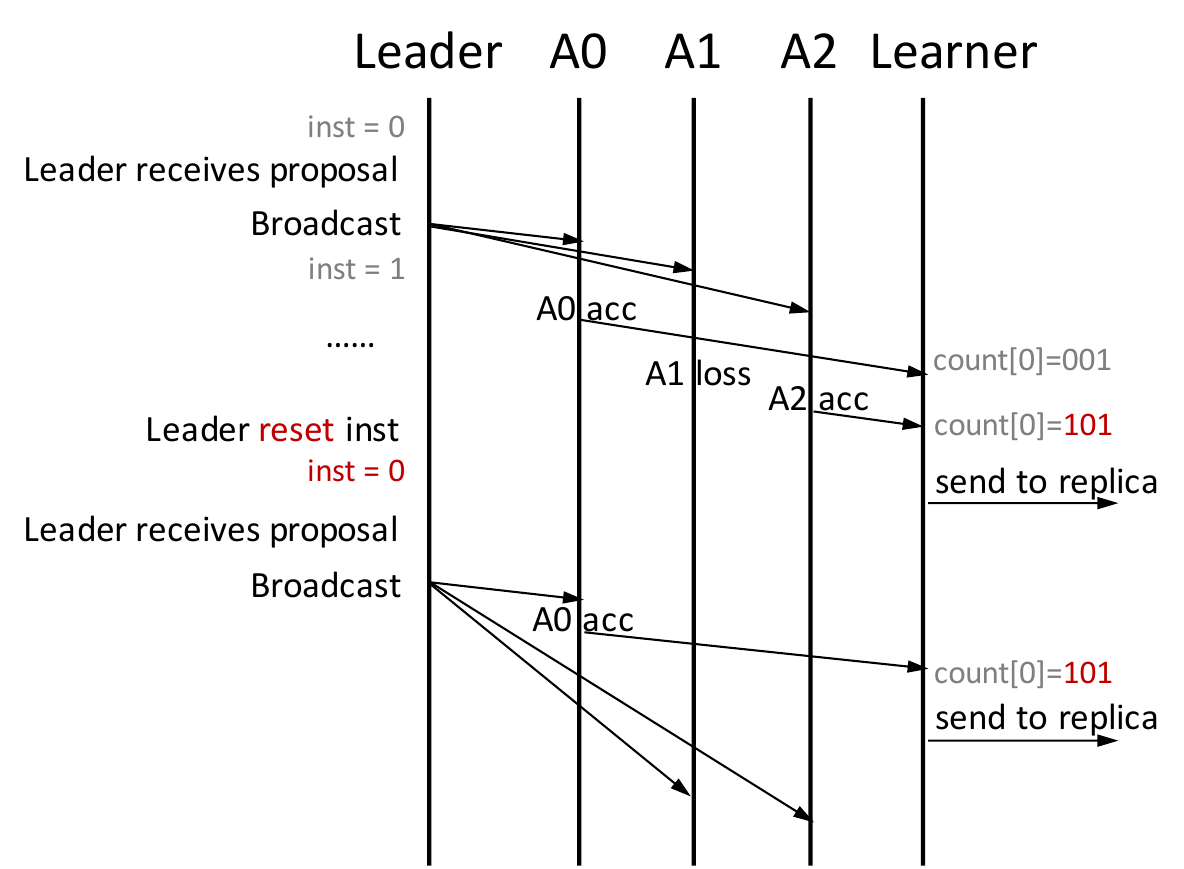}
\caption{Timing diagram of majority-quorum violation in P4xos. \textnormal{Inst represents the instance number; A0, A1, A2 are the three acceptors; count is a register array where each slot tracks the number of acceptors that have accepted the value for a given instance.}} \label{fig:p4xos-result}
\vspace{-5mm}
\end{figure}
\begin{table*}[t]
\centering
\resizebox{0.8\textwidth}{!}{
\begin{tabular}{|l|l|c|c|c|c|}
\hline

\multirow{2}{*}{\textbf{Benchmark}} & \multirow{2}{*}{\textbf{Bug Description}} & \multicolumn{2}{c|}{\textbf{\system}} & \multicolumn{2}{c|}{\textbf{p4tv}} \\
\cline{3-6}
 &  & \textbf{Time (s)} & \textbf{Mem (MB)} & \textbf{Time (s)} & \textbf{Mem (MB)} \\
\hline

\multirow{2}{*}{\textbf{FRR}~\cite{frr}} 
& Bug 1: Unexpected Mirror & 2.4 & 262 & 43.57 & 352 \\
& Bug 2: State Inconsistency  & 2.1 & 260 & 67.14 & 325 \\
\hline

\multirow{2}{*}{\textbf{P4NIS}~\cite{p4nis}} 
& Bug 1: Forwarding Sequence Desync  & 0.17 & 178  & 58.56 & 309 \\
& Bug 2: Tunnel State Leakage & 0.1 & 140 & 59.07 & 306  \\
\hline

\textbf{DDOSD}~\cite{ddosd} & Vuln: Window Label Collision & 0.12 & 141 & 109.7 & 348 \\
\hline

\textbf{Cheetah}~\cite{cheetah} & Bug: Slot Index Collision  & 0.22 & 137  & 160.6 & 396 \\
\hline

\end{tabular}
}
\caption{Comparative results of bug finding across different P4 verification tools.}
\vspace{-5mm}
\label{tab:comparison}
\end{table*}
\para{P4xos.} 
P4xos implements the Paxos consensus protocol directly in programmable switches, with roles (leader, acceptor, learner) executed in the data plane. 
The leader manages instance numbers, while acceptors and learners use them as register indices to track consensus rounds. 
The learner counts votes and forwards values to replicas once a majority quorum is reached.

\system uncovered three bugs in P4xos. 
Two of them are local implementation errors within individual roles that had already been reported by prior work \cite{p4v,vera}: 
one caused by register out-of-bounds access, and another due to the drop\_flag not being correctly set. 
The third bug, however, is a previously unknown violation of the majority-quorum property, which spans multiple distributed nodes. 
Since this property is fundamental to consensus algorithms, its violation breaks consistency guarantees across nodes. 
After fixing the two known bugs, \system revealed that the system as a whole still violated the majority-quorum property. 
We now illustrate the root cause of this bug.

Specifically, the learner utilizes registers to maintain vote counts, but lacks a mechanism to clear register values after each instance.
Residual vote counts persist across instances, so when the leader resets or wraps the instance number (inst), the learner may reuse stale values.
Figure \ref{fig:p4xos-result} presents a timing diagram illustrating the bug. 
The system works correctly for inst = 0, but after several rounds, the inst is either wrapped around or explicitly reset by the leader.
At this point, the learner inherits the outdated quorum state (101).
With only one new vote, it may falsely detect a majority and prematurely forward a value, violating the majority-quorum requirement and leading to consensus incorrectness.

\para{DistCache.}
We discovered an integer wrap-around bug in DistCache, a multi-switch caching system adapted from FarReach~\cite{sheng2023farreach}. 
DistCache uses a ``power-of-two-choices'' scheme to balance load across leaf and spine caches. 
However, \system reveals that the integer counters (\texttt{leafload}, \texttt{spineload}) tracking cache load can overflow under sustained traffic, wrapping to small or negative values.

When this occurs, the proxy switch (\emph{clienttrack}) mistakenly interprets the wrapped value as low load and continues directing traffic to the already overloaded leaf cache, 
leaving the spine underutilized. The imbalance persists until the counter increments wrap again, prolonging recovery. 
The bug stems from the lack of explicit overflow handling in load counters, leading to severe traffic skew, low overall throughput, and high query latency.

\para{Gecko.}
\label{sec:gecko-bugs}
%We have found three bugs in Gecko. These three bugs were detected first by Procurator. Before detailing these discoveries, we first outline Gecko's architecture and key design choices.
Gecko is a system that leverages programmable P4 switches to reduce the tail latency of distributed key-value stores. 
Its topology consists of a Master, a Client, a single P4 switch (with two pipelines), and three chunk nodes. A write request proceeds as follows: the Client sends the request to the one pipeline of the swtich, which forwards it to the Master. The Master responds to the switch, and the other pipeline then replicates the request to all three chunk servers. Once any two servers acknowledge completion, the switch immediately notifies the Client. A core feature of Gecko is its recirculating \emph{timer} packet, which maintains a global time register to support in-switch timeout detection and ``chasing'' functionality.

\system identified three bugs in Gecko:

{\it Timer Packet Loss or Blockage.} If the timer packet is dropped or blocked at the boundary between the switch and the external network, the time register stalls, disabling timeout detection. Pending write requests in registers remain unevaluated, causing inconsistencies and indefinite waiting.

{\it Limited Concurrency Handling.} If the Master injects multiple write requests into the first pipeline in quick succession, Gecko’s timer logic updates only one request’s timeout state per pipeline traversal. Under high concurrency, the timer mechanism
struggles to keep pace with incoming requests, leading to inaccurate or delayed timeout detections.

{\it Improper Timer Initialization.} If a write request arrives before the timer circuit is active, it bypasses in-switch logic and falls back to Master-handled coordination. This breaks Gecko’s low-latency design and may cause inreliability.

\begin{table*}
\centering
\resizebox{0.60\textwidth}{!}{
\begin{tabular}{|l|c|c|c|c|c|c|}
\hline
\multirow{2}{*}{\textbf{System}} & \multicolumn{3}{c|}{\textbf{without slicing}} & \multicolumn{3}{c|}{\textbf{with slicing}} \\
\cline{2-7} 
 & \textbf{States} & \textbf{Memory} & \textbf{Time} & \textbf{States} & \textbf{Memory} & \textbf{Time} \\
\hline
\textbf{Netchain} & 229374 & 17.45GB & 1min39s & 229374 & 7.05GB & 50s \\
\textbf{P4xos} & 10765 & 17.81GB & 1min27s & 10765 & 13.86GB & 1min7s \\
\textbf{Gecko} & 1348 & 9.8GB & 45s & 1348 & 1.26GB & 4.7s \\
ATP & 72 & 131MB & 0.17s & 72 & 40MB & 0.07s \\
\textbf{DistCach} & 163858 & 2.98GB & 18.5s & 163858 & 2.91GB & 18s \\
\hline
\end{tabular}
}
\caption{Results showing the impact of slicing on memory and execution time.}
\vspace{-3mm}
\label{tab:ablation}
\end{table*}
\para{ATP.}
ATP provides \emph{best-effort} in-network aggregation for distributed training (DT) jobs \cite{atp}. It allocates a small portion of programmable switch memory for ``aggregators'', indexed by a hash of the \texttt{(JobID, SequenceNumber)} pair. Workers send gradient fragments to ToR switches, which partially aggregate them before forwarding results to the parameter server. ATP includes mechanisms for collision detection (\eg hash conflicts), fallback to end-host aggregation, and reliability features such as retransmissions on out-of-order acknowledgments. We analyze an open-source P4\_16 implementation and \system uncovers two issues:

{\it Indefinite Aggregation via 8-bit \texttt{agtr\_time}.} ATP uses an 8-bit field (\texttt{p4ml.agtr\_time}) to track aggregation progress, with a threshold (\eg, 32). If the field exceeds this threshold (\eg set to 40), the switch continually waits for a completion condition that never occurs. Aggregation is left unfinished, and no response is returned. In our experiments, a host transmitted 40 packets, all of which were silently aggregated but never finalized.

{\it Mismatch under Packet Loss.} With packet loss, the switch’s count of completed aggregations sometimes diverges from the actual number of aggregation requests issued. While ATP’s best-effort semantics may tolerate partial aggregation for ML workloads, formally this violates correctness guarantees. In controlled tests with 1000 packets and induced losses, \system found states where recorded completions did not match the sender’s intended updates, revealing a latent violation of delivery and aggregation guarantees.

\subsection{Verification on Single-Switch Programs}\label{sec:comparison}

We next evaluate whether \system offers advantages over the state-of-the-art stateful P4 verifier, p4tv~\cite{p4tv}. 
We focus on p4tv because it focusing on the task of verifying Linear Temporal Logic properties, and generates execution traces upon verification failures, allowing us to confirm that the reported errors match the bugs we identified. Other verifiers are either stateless~\cite{bf4, p4v} or lack trace support and LTL capability ~\cite{p4inv}, which hinders root cause analysis.
Since p4tv only supports single-device in-network programs, we conduct experiments on representative ones, including FRR~\cite{frr}, P4NIS~\cite{p4nis}, Cheetah~\cite{cheetah}, and DDOSD~\cite{ddosd}.

%In DDOSD, \system successfully detected a \textit{Window Label Collision} where an 8-bit counter overflow corrupts the statistical model. For Cheetah, it relvealed a \textit{Slot Index Collision} that causes active connections to be overwritten. In P4NIS, it identified two bugs: a forwarding sequence desynchronization from a counter wrap-around and a \textit{Tunnel State Leakage} violating traffic isolation. Finally, in the \textbf{FRR} fast re-route program, we checked for erroneous packet mirroring and a \textit{BFS State Inconsistency} that leads to incorrect path computations.
%\qs{what is bit-precise modeling?}

%In terms of bug detection, the results in Table~\ref{tab:comparison} show that \system~achieves both equivalence and superiority. On FRR, P4NIS, and Cheetah, we successfully identified the same set of bugs as p4tv. On ddosd, \system~uncover a subtle bug that p4tv failed to detect. \system~'s ability to find the additional bug in ddosd stems from the higher fidelity of its modeling approach. Specifically, \system~found a Window Label Collision vulnerability in the ddosd system, which uses a count sketch to detect DDoS attacks. This vulnerability arises because an 8-bit window label used to tag statistical entries overflows after reaching 255 and wraps around to 0. This is a security vulnerability that stems from the fixed-width integer constraints of the hardware. \system~'s bit-precise modeling approach successfully detect this bug, while p4tv uses a higher-level integer abstraction, failed to find it. 

Table~\ref{tab:comparison} reports the detected bugs, verification time, and memory consumption of \system and p4tv. Both tools detect the same set of bugs across all in-network systems, but \system achieves significantly better performance. 
Specifically, compared to p4tv, \system efficiently reduces the memory consumption by 1.9$\times$ on Cheetah, and reduces the verification time by 913.2$\times$ on DDOSD.
By leveraging the Semantic-Aware State Pruner, \system generates only the packet sequences necessary to expose a bug, effectively pruning the state space. In contrast, p4tv models incoming packets as non-deterministic within an infinite loop, forcing its SMT-based model checker to exhaustively explore a vast space of packet inputs and interleavings. This exhaustive exploration is computationally expensive, particularly for bugs that require long and specific state-transition sequences.

%In terms of performance, \system~'s methodology offers significant advantages over p4tv. Procurator uses a lightweight, user-defined host model to generate the exact packet sequences required to trigger a bug. This guided approach effectively prunes the state space that needs to be explored. In contrast, p4tv's model treats incoming packets as non-deterministic within an infinite loop, requiring its SMT-based model checker to exhaustively search a vast state space of possible packet inputs and program interleavings. This is computationally expensive, especially for bugs that depend on a specific, long sequence of state transitions.

%In conclusion, these results demonstrate that \system~not only introduces the capability to verify distributed P4 systems but also naturally subsumes single-device verification, where its high-fidelity model enables it to uncover subtle bugs missed by existing state-of-the-art tools.

% \vspace{-1em}
\subsection{Overhead}\label{sec:exp-slicing}

We now examine \system’s overhead by measuring memory usage and execution time while verifying six real-world systems. Table~\ref{tab:ablation} compares results with and without the pruning optimization. The results show that pruning significantly reduces both memory consumption and verification time, highlighting its effectiveness in eliminating redundant or unnecessary components from the system’s state representation. We next analyze \system’s verification overhead on individual systems.

%To determine the effectiveness of the Procurator's pruning algorithm, we conducted an ablation study to evaluate its impact on efficiency and resource utilization. The results, presented in Table \ref{tab:ablation}, demonstrate that applying pruning not only lowers the overall memory consumption but also decreases the processing overhead, resulting in faster execution time. 

%The primary reason for these improvements is that pruning eliminates redundant or unnecessary components from the system's state representation. The effectiveness of these reductions depends largely on how relevant the property is to the system's components. 

%\para{Memory consumption reduction.} Systems like Gecko show exceptional improvements in memory utilization, with a dramatic reduction of 87\% due to the significant elimination of irrelevant data. This is primarily because Gecko contains many components that are not critical to the specific properties being verified, making pruning highly effective. In contrast, systems like DistCache show minimal reductions, indicating that much of the data is essential to the system's operation and, thus less amenable to optimization through pruning. For most other systems, the reduction in memory usage falls somewhere in between. Netchain and P4xos, for example, experience moderate reductions, with Netchain reducing its memory usage by 60\% and P4xos reducing by 22\%. These results suggest that for systems with a mix of relevant and irrelevant components, pruning can still lead to significant, though not extreme, improvements in memory efficiency.

The pruning optimization achieves the largest improvements for Gecko, reducing memory usage by 87\% and execution time by 90\% by removing components irrelevant to the verified properties. This is primarily because Gecko contains many components that are not critical to the specific properties being verified, making pruning highly effective. In contrast, DistCache sees minimal memory reduction, as most of its states are essential. Netchain and P4xos experience moderate gains, with memory reductions of 60\% and 22\% and execution time reductions of 50\% and 23\%, respectively. These results indicate that pruning can substantially enhance memory and time efficiency for systems containing a mix of relevant and irrelevant components, though the benefit depends on the proportion of non-essential data.

\section{Related Work}\label{sec:relatedwork}

We discuss related work to \system in this section.

%\begin{sloppypar}
\noindent{\bf P4 program verification.} 
P4-NoD~\cite{p4nod}, ASSERT-P4~\cite{assert-p4-sosr, assert-p4-conext}, and Vera\cite{vera} translate P4 into other languages (\eg SEFL) and use \textit{symbolic execution} to explore execution paths and check assertions. p4v~\cite{p4v}, bf4~\cite{bf4}, and Aquila~\cite{aquila} improve scalability via \textit{deductive verification}, formally reasoning about weakest pre-conditions to prove correctness. P4Inv~\cite{p4inv} partially addresses this by modeling programs as infinite loops and inferring packet invariants. In contrast, \system models P4 programs using the actor paradigm and CSP models, capturing infinite-loop behavior, supporting features like recirculation and mirroring, and enabling verification of complex interactions.

%P4-NoD\cite{p4nod}, ASSERT-P4\cite{assert-p4-sosr, assert-p4-conext}, and Vera\cite{vera} translate P4 into other language (e.g., SEFL) and utilize \textbf{Symbolic Execution} tools to explore execution paths and check the satisfaction of assertions. p4v\cite{p4v}, bf4\cite{bf4}, and Aquila\cite{aquila} further enhance the scalability of verification by employing \textbf{Deductive Verification} techniques, which reason about the weakest preconditions of a program through formal mathematical proof methods to prove its correctness. Capisce\cite{capisce} computes a control interface specification (ci-spec) for a P4 program, ensuring that control plane configurations satisfying the ci-spec will not cause the data plane program to violate its properties. However, these approaches do not fully account for the correct modeling of stateful data (e.g., registers), which is critical for accurately verifying programs that involve state-dependent logic. P4Inv\cite{p4inv} takes a first step in this area by modeling the P4 program as an infinite loop (i.e., while(true)), automatically inferring packet invariants to ensure that the assertion is guaranteed in a given configuration. \system~models P4 programs as process models with channels: on the one hand, it naturally captures the infinite loop characteristic of switches while additionally supporting P4 features such as recirculate and mirror; on the other hand, it can models complex interactions between different P4 programs, thus enabling the verification of distributed P4 systems.

\noindent{\bf Stateful network verification.} 
Stateful network verification is crucial for ensuring the correctness and reliability of modern networks with stateful components like middleware. SymNet~\cite{symnet} uses symbolic execution with packet-header state and introduces SEFL for efficiency. VMN~\cite{vmn} employs SMT solvers and abstracts middleware to reduce network scale and speed up verification. Velner et al.~\cite{velner2016some} made region traffic isolation verification decidable via packet-order abstraction, later optimized by Alpernas et al.~\cite{alpernas2018abstract}. NetSMC\cite{netsmc} simplifies the network to a single packet model and improves scalability with a custom symbolic model checking algorithm. In contrast, \system targets on capturing the internal interactive behaviors in real-world distributed in-network programs.

%Stateful network verification is a vital area of research for ensuring the correctness and reliability of modern network systems, especially those using stateful components like middleware. SymNet\cite{symnet} leverages symbolic execution with state information in packet headers and introduces SEFL for efficiency. VMN\cite{vmn} adopts SMT solvers, proposing an abstract middleware modeling technique that improves verification speed by reducing the network scale. Velner et al.\cite{velner2016some} made region traffic isolation verification decidable through packet processing order abstraction, while Alpernas et al.\cite{alpernas2018abstract} further optimized this approach. NetSMC\cite{netsmc} simplifies the network to a single packet model and improves scalability with a custom symbolic model checking algorithm. Our system focuses not only on the distributed in-network computing programs but also on the impact of interleaved executions in distributed P4 systems on the state.

%\para{Model checking.} Previous work has applied model checking to network verification. NICE\cite{nice} verifies OpenFlow applications by checking network invariants. Kothari et al.\cite{musuvathi2004model} use abstractions to manage state space explosion and ensure protocol correctness in large-scale systems. Plankton\cite{plankton} improves network configuration verification by combining equivalence partitioning with explicit-state model checking. To the best of our knowledge, \system~is the first approach to apply model checking for the verification of distributed P4 systems.

% \vspace{-1em}
\section{Conclusion}\label{sec:conclusion}

This paper presents \system, an efficient verification framework for distributed in-network programs. By modeling pipelines as reactive actors, unifying their interactions as message passing within CSP, and supporting flexible property specification through a unified intent language, \system effectively captures interactive behaviors and detects subtle related bugs. To ensure scalability, it incorporates a semantic-aware state pruner that reduces verification complexity. Evaluation results show that \system not only uncovers previously overlooked bugs in real-world systems but also achieves a significant reduction in verification time and memory consumption.

\newpage

\bibliographystyle{plain}
\bibliography{reference}

\clearpage
\appendix

\appendix
\section*{Appendices}

\section{Abstracting Parsing and Deparsing}
\label{app:pardepar}

%To avoid modeling the complex, bit-level data flow within P4 parsers and deparsers, we employ an efficient abstraction. 
We model parsing and deparsing as atomic operations on the validity of headers, represented by the ``.valid'' flag associated with each header instance.

\subsection{Parser Semantics}
%For the parser, every time its state machine successfully extracts a header from the packet stream, we model this event by setting the `.valid` flag of the corresponding header to \texttt{true}. This abstraction preserves the essential control-flow information that the program logic depends on, without needing to model the actual bit-level extraction.

For the parser, each successful header extraction is modeled by setting the corresponding ``.valid'' flag to \texttt{true}. We formalize the parser as a function $\text{Parse}(p_{raw}) = p$, transforming a raw packet into a structured one with header validity flags. A header’s validity is set to \texttt{true} if the execution path reaches a parser state with an \texttt{extract} statement.

%We model the overall behavior of the parser as a function $\text{Parse}(p_{raw}) = p$, which transforms a raw packet into a structured packet with header validity flags. Its core semantics are defined by the following rule, which ties header validity to the execution path reaching a parser state that contains an `extract` statement.

% \begin{center}
% \text{[Parse-Header]}
% \end{center}
\begin{equation}
\frac{
   p_{raw} \rightarrow_{\mathcal{P}} s \quad \land \quad \texttt{extract}(h_i) \in \text{body}(s)
}{
    (\text{Parse}(p_{raw})).h_i.\text{valid} = \text{true}
}
\end{equation}
Here, $p_{raw} \rightarrow_{\mathcal{P}} s$ denotes that the processing path for packet $p_{raw}$ in parser $\mathcal{P}$ includes the state $s$, and $\texttt{extract}(h_i) \in \text{body}(s)$ denotes that an `extract` statement for header $h_i$ is syntactically present in the definition of state $s$. 
%This rule abstracts away bit-level details but precisely retains the core information of "whether a header exists," which is critical for the subsequent ingress control flow.

\subsection{Deparser Semantics}
%For the deparser, we precisely simulate the behavior of \texttt{emit()} calls. Within the deparser control block, each call to \texttt{emit()} is modeled as an update to a temporary set of headers intended for emission. Before the packet is serialized, a final, architectural cleanup step inspects this set and resets (i.e., sets `.valid` to \texttt{false}) any header that was not explicitly marked for emission.

For the deparser, each \texttt{emit()} call is modeled as adding the corresponding header to a temporary emission set. Before serialization, a final cleanup step resets the ``.valid'' flag of any header not in this set, ensuring only explicitly emitted headers remain valid.

We model the deparser's behavior as a two-phase function, $\text{Deparse}(p) = p_{final}$.

\noindent{\bf Phase 1: Deparser Control Block Execution.}
This phase models the effect of \texttt{emit()} calls. We introduce a temporary set, \texttt{EmitSet}, which is constructed based on the deparser's execution trace.

\begin{equation}
\frac{
   \texttt{emit(p.h\_i)} \in \text{trace}(\text{exec}(\mathcal{D}, p))
}{
   h_i \in \text{EmitSet}
}
\end{equation}

In this rule, $\text{trace}(\text{exec}(\mathcal{D}, p))$ represents the sequence of statements that are actually executed when the deparser control block $\mathcal{D}$ is run on packet $p$. The premise holds if an $\texttt{emit}$ call for header $h_i$ is part of this trace, correctly handling conditional logic. This translates the imperative action of an \texttt{emit()} call into membership in the \texttt{EmitSet}.

\noindent{\bf Phase 2: Architectural Cleanup.}
After all logic in the deparser control block has been executed, a final cleanup step determines the final validity state of each header based on the contents of \texttt{EmitSet}. Let $\mathcal{H}_p$ be the set of all header instances defined for packet $p$.

\begin{equation}
\frac{
   h_j \in \mathcal{H}_p \land h_j \notin \text{EmitSet}
}{
   p_{final}.h_j.\text{valid} = \text{false}
}
\end{equation}

This rule precisely models the P4 architectural specification: only headers that are explicitly emitted are preserved in the output packet. Any header that was valid before the deparser but was not emitted is implicitly dropped (i.e., its `.valid` bit is cleared). This mechanism is crucial for verifying programs that rely on dynamically modifying packet headers.

\section{Additional TM Semantic}
\label{app:tmse}

\begin{equation}
\frac{
    p.\text{meta.clone\_spec} = \text{I2E} \quad p_{\mathrm{clone}} = \text{clone}(p)
}{
    \mathrm{TM}(p) = \mathrm{enq}(\mathrm{enq}([\ ], p),\ p_{\mathrm{clone}})
}
\end{equation}
This rule models the ingress-to-egress mirror primitive. The TM duplicates the packet and enqueues both the original and the clone into the egress channel. The final disposition of both packets will be determined during the egress-draining phase of the current actor reaction.

\begin{equation}
\frac{
    p.\text{meta.clone\_spec} = \text{I2I} \quad p'_{\mathrm{clone}} = \text{clone\_and\_mark\_i2i}(p)
}{
    \mathrm{TM}(p) = \mathrm{enq}(\mathrm{enq}([\ ], p),\ p'_{\mathrm{clone}})
}
\end{equation}
This rule models the first step of an ingress-to-ingress clone. The TM duplicates the packet but applies a special mark to the clone. This mark serves as an internal message for the \texttt{FinalDispatch} stage. The marked packet transits the egress pipeline merely as a token to unify the model, instructing FD to generate a `recirc` event on its behalf.

\begin{equation}
\frac{
    p.\text{meta.drop\_flag} = \text{true}
}{
    \mathrm{TM}(p) = [\ ]
}
\end{equation}
This rule models packet drops within the ingress pipeline. If a packet is marked to be dropped, the TM produces an empty egress queue, effectively terminating the packet's processing journey before it can enter the egress pipeline.

\section{Additional FD Semantic}
\label{app:fdse}

\begin{equation}
\frac{
    p.\text{meta.is\_i2i\_clone} = \text{true}
}{
    \mathrm{FD}(p) = \big(\ [\ (I_{\mathrm{self}},\ p')\ ],\ [\ ]\ \big)
}
\end{equation}
This rule completes the I2I mirror operation by interpreting the internal message from \textbf{[TM-I2I-Clone]}. While architecturally the packet bypasses egress, our model centralizes all disposition logic here. This rule sees the marked packet and converts it into a \texttt{recirc} event, correctly modeling the end-to-end behavior of a packet re-entering the ingress pipeline.

\begin{equation}
\frac{
    p.\text{meta.drop\_flag} = \text{true}
}{
    \mathrm{FD}(p) = \big(\ [\ ],\ [\ ]\ \big)
}
\end{equation}
This rule models packet drops within the egress pipeline. If a packet is marked to be dropped at this stage, FD produces an empty set of events. The packet's processing is silently terminated with no observable external effect.

\section{Executable Translation for Model Checking}
\label{app:translation-appendix}

\para{Modeling assumptions and non-blocking guarantees.}
Our reaction semantics requires that a single pipeline pass (\emph{Ingress-triggered, Egress-draining}) executes atomically without internal interleavings. To enforce this in Promela, we adopt the following modeling constraints for the executable model:

(1) \emph{FIFO and bounded channels.} Each device's ingress channel and each device-local egress queue are FIFO and bounded. We choose per-device bounds \texttt{Q\_IN} and \texttt{Q\_EG} large enough so that all \texttt{!} operations inside a single \texttt{atomic\{\}} do not block (i.e., the egress queue does not overflow; \texttt{FinalDispatch} never blocks on an enqueue to an ingress). Violations trigger explicit \texttt{assert}s.

(2) \emph{Local egress queue implements Drain.} Egress draining (Rule (~\ref{eq:drain-empty})(~\ref{eq:drain-step})) is realized by a device-local queue \texttt{egress\_q} that is fully emptied before the reaction ends.

(3) \emph{FD is the sole observable exit.} All observable events (our external observations) are produced only by \texttt{FinalDispatch} as \texttt{!} into some ingress channel, matching the observable alphabet in \S\ref{sec:formal-model}.

(4) \emph{Links resolve forwarding deterministically.} The mapping from \texttt{(self, egress\_spec)} to the destination's ingress channel is fixed by the network topology \textsf{Links}; if no mapping exists, the packet is dropped at FD (producing no observable event).
\paragraph{Global bounds and packet skeleton.}
We expose bounds as preprocessor constants and use a minimal packet skeleton that carries only fields read by the translation (clone/recirc/drop flags and egress port). These names correspond to the meta fields referenced by $\Pi_I/\Pi_E$ and FD in the core semantics.

{%
\fontsize{10pt}{12pt}\selectfont
\begin{verbatim}
#define Q_IN  8    
#define Q_EG  8    

mtype = { NO_CLONE, I2E, I2I };
typedef packet_t {
  mtype clone_spec;
  bool  drop_flag;       
  bool  recirc_flag;     
  bool  is_i2i_clone;    
  byte  egress_spec;     
}
\end{verbatim}
}

\paragraph{Links resolving function.}
We encode \textsf{Links} as a guarded \texttt{send\_to} helper. This is the only place where a device may send to another device's ingress.
{%
\fontsize{10pt}{12pt}\selectfont
\begin{verbatim}
inline send_to(byte self, byte egress_spec
, packet_t p)
{
    if
    :: self == 1 && egress_spec == 10 -> 
    s2_ingress ! p
    :: self == 2 && egress_spec == 20 -> 
    s1_ingress ! p
    :: else -> skip     
    fi
}
\end{verbatim}
}

\paragraph{Pipeline stubs.}
The following stubs stand for the (inlined) realizations of $\Pi_I$/$\Pi_E$ over parser/control/deparser. The translation emits concrete bodies from the P4 IR; here we present their signatures.
{%
\fontsize{10pt}{12pt}\selectfont
\begin{verbatim}
inline IngressParser(packet_t p)   {}
inline Ingress(packet_t p)         {}
inline IngressDeparser(packet_t p) {}
inline EgressParser(packet_t p)    {}
inline Egress(packet_t p)          {}
inline EgressDeparser(packet_t p)  {}
inline get_mirror_port(packet_t p) {}
\end{verbatim}
}
\paragraph{Faithfulness to the abstract semantics.}
By construction: (i) a single reaction is executed as one \textit{atomic\{\}} block (\emph{Ingress-triggered, Egress-draining}); (ii) Drain is realized by the local \textit{egress\_q} loop with the concatenation discipline $L=L_p \cdot L_0$; (iii) all observable effects are \textit{!} into ingress channels at \textit{FinalDispatch}; and (iv) forwarding targets are determined solely by \textsf{Links}. Together with the non-blocking bounds \textit{Q\_IN}/\textit{Q\_EG} and the \textit{assert}s above, this executable model is a faithful instance of the formal semantics used in our soundness argument.

\end{document}